# FORESEE: A Cooperative Lane Change Model for Connected and Automated Driving

Rafael Molina-Masegosa[a], Sergei S. Avedisov[b], Miguel Sepulcre[a],
Javier Gozalvez[a], Yashar Z. Farid[b], Onur Altintas[b]
[a] Universidad Miguel Hernandez de Elche, Elche, Spain, {rafael.molinam, msepulcre, j.gozalvez}@umh.es
[b] Toyota Motor North America R&D – InfoTech Labs, Mountain View, CA, USA, {sergei.avedisov, yashar.zeyinali.farid, onur.altintas}@toyota.com

***Abstract*— This paper presents FORESEE, a novel cooperative lane change model for connected and automated driving. FORESEE leverages Vehicle-to-Everything (V2X) data to anticipate traffic conditions and effectively organize lane changes. Specifically, it uses V2X data to organize vehicles into lanes based on their desired speeds, which helps to homogenize traffic flow and reduce disturbances caused by speed differences among vehicles within the same lane. The study demonstrates that implementing cooperative lane changes with FORESEE enhances average vehicle speed and energy efficiency compared to non-cooperative lane changes, which typically rely on short-term and local information about the ego vehicle and its immediate neighbors. This is achieved through fewer but more effective lane changes. Additionally, vehicles can maintain speeds closer to their desired speeds, resulting in fewer fluctuations in speed and acceleration and enhanced driving comfort. Moreover, cooperative lane changes can better manage road traffic disturbances, such as obstacles, by anticipating traffic conditions and organizing lane changes ahead. FORESEE serves as a valuable framework for the future design and testing of V2X-based maneuver coordinations as their effectiveness depends on how vehicles change lanes and their ability to plan and organize maneuvers in consideration of the upcoming traffic conditions.**

## I. Introduction

Connected and Automated Vehicles (CAVs) utilize Vehicle-to-Everything (V2X) communications to exchange information and extend their field of view, providing a more accurate and comprehensive understanding of the driving environment. This augmented perspective can enhance both driving safety and traffic efficiency through cooperative mobility solutions. A key example is maneuver coordination, also known as maneuver sharing and coordination or cooperative driving. With maneuver coordination, vehicles do not need to infer the intentions of nearby vehicles. Instead, they directly exchange information about their driving intentions and coordinate their maneuvers using V2X messages. Maneuver coordinations can positively impact the traffic efficiency if vehicles can plan and organize their maneuvers considering the driving conditions ahead, which can be inferred from V2X data. The effectiveness of cooperative mobility solutions, including maneuver coordination, is closely tied to how vehicles change lanes. Lane changes are critical as they significantly impact mobility, traffic efficiency, and the performance of the ego vehicle [1].

Existing lane change schemes are primarily non-cooperative, meaning that vehicles rely only on their own sensing and local observations. As a result, interactions remain implicit and the decision process can only consider the effect on the ego vehicle and its immediate neighbors. This approach limits the capacity to plan and organize lane changes for overall traffic efficiency. For example, a vehicle using a non-cooperative lane change scheme may switch lanes for an immediate acceleration advantage, only to lose this benefit if a slower vehicle is present ahead in the new lane. To overcome these limitations, this study introduces a cooperative lane change model in which vehicles share context information through V2X. This explicit communication allows them to make lane change decisions that are more consistent with traffic efficiency. This concept of cooperative lane change refers to lane changes that are triggered thanks to a V2X-enhanced decision-making, and should not be confused with the concept of coordinated lane change, which require two or more vehicles to negotiate and execute a coordinated maneuver.

The proposed model, FORESEE (look-ahead lane change for speed-based vehicle ordering), is a decentralized cooperative lane-change model that leverages V2X data to anticipate traffic conditions ahead and evaluates lane-change decisions considering both short-term and long-term driving effects to improve system-level traffic efficiency. Specifically, FORESEE uses V2X data to estimate the speed of each lane and arranges vehicles into lanes according to their desired speed. The desired speed of a vehicle is defined as the speed it would maintain under free-flow conditions on an empty road[1]. It is a vehicle-specific parameter that reflects its preferred cruising behavior. In congested or constrained traffic conditions, vehicles are typically unable to reach their desired speed and instead must adjust to an actual speed dictated by the behavior of preceding vehicles and the traffic density. Our approach helps homogenize traffic flow and reduce disturbances often caused by abrupt lane changes or braking maneuvers triggered by speed differences between vehicles in the same lane. The study demonstrates that cooperative lane changes using FORESEE enhance average vehicle speed and driving efficiency compared to non-cooperative lane changes based solely on local information.

[1] The desired speed is also known as free flow speed. It is an important parameter of the most common car following models like IDM [2], OVM [3] or FVDM [4]. The desired speed is not necessarily the same for all vehicles.

.

FORESEE achieves this with fewer but more effective lane changes by strategically organizing lane changes anticipating the road traffic conditions ahead using V2X data. This allows vehicles to travel closer to their desired speeds, reducing speed variability and accelerations, which in turn enhances driving comfort. Furthermore, the study shows that cooperative lane changes can better handle road traffic disturbances by anticipating their impact and organizing lane changes to reduce the number of vehicles affected, thereby further increasing vehicle speeds. FORESEE also provides a natural basis for coordinated maneuvers, as it already integrates look-ahead V2X information and aligns lane change decisions with the overall traffic context. This makes it compatible with multi-vehicle maneuver coordination mechanisms when a desired lane change requires cooperation.

The rest of the paper is organized as follows. Section II introduces the state of the art on cooperative and non-cooperative lane-change models and situates our cooperative model in relation to existing approaches. Section III analyzes traffic dynamics and efficiency using a non-cooperative lane change model, and explores how traffic behaves when vehicles do not leverage V2X data for their lane change decisions. Section IV introduces FORESEE, our proposed cooperative lane change model, and illustrates the impact of both non-cooperative and cooperative lane changes across three representative scenarios. Section V evaluates the impact of cooperative lane changes on traffic dynamics and efficiency using the FORESEE model. Section VI highlights and discusses the potential benefits of maneuver coordination. Finally, Section VII presents the conclusions of our study.

## II. State of the Art

Lane changes are designed at two levels [5]. The control level focuses on the physical execution of lane changes and analyzes steering wheel input, throttle, and brake control to derive a smooth and effective trajectory. The strategy level [5] decides when to perform a lane change. Most cooperative lane change models found in the literature operate at the control level and use V2X information to support trajectory planning and execution [6][7][8]. In contrast, strategy-level lane change models, determine when a lane change should be initiated. The work presented in this paper, and the proposed FORESEE model, focus on the strategy level. Most existing strategy level lane change models decide when to execute a lane change based only on local information, typically limited to the ego vehicle and the vehicles immediately behind in the current and target lanes. Among these non-cooperative approaches, the MOBIL model [9] is one of the most widely used in both human and automated driving studies ([5], [10], [11], [12]). MOBIL focuses on minimizing the immediate deceleration of vehicles directly affected by a lane change and is frequently adopted as a baseline in lane change research.

A limited number of studies have exploited V2X communications at the strategy level. Existing works differ significantly in their objectives, architectural assumptions, and operational scope, and therefore address problems that are only partially related to general-purpose lane-change decision making. An extensive review in [10] highlights that most cooperative approaches have primarily focused on local, individual or vehicle-centric objectives rather than explicitly targeting traffic efficiency at the system level. Furthermore, many evaluations were conducted under free-flow conditions, where lane changes are not critical to traffic performance, limiting the ability to assess their impact on realistic and heterogeneous traffic scenarios. These observations remain largely valid for current cooperative strategy-level approaches.

Some studies pursue system-level objectives but rely on centralized or infrastructure-supported decisions. For example, [13] proposes a lane assignment mechanism in which a roadside controller determines the appropriate lane for each vehicle and communicates these decisions via V2X. While effective under ideal deployment conditions, such approaches assume centralized control and continuous infrastructure support, which differs fundamentally from decentralized decision-making paradigms in which vehicles act autonomously based on locally available information.

Other works extend existing decentralized strategy-level lane-change models by incorporating additional information or modified incentive formulations while preserving ego-centered decision criteria. The framework proposed in [14] combines MOBIL with a modified FVDM car-following model and assumes access to upstream information from connected vehicles. However, because the incentive structure continues to rely on instantaneous acceleration gains inherited from MOBIL, the resulting decisions mainly reflect short-term local benefits and cannot fully exploit extended look-ahead information. The work in [12] further augments MOBIL with an additional incentive anticipating mandatory lane changes, enabling vehicles to progressively move toward required route lanes. While these approaches incorporate cooperative information, their objectives remain centered on individual vehicle performance rather than continuous traffic-level organization.

Most of the relevant decentralized strategy-level lane-change models identified are designed for specific traffic situations. [12] is designed for incident-driven evasive maneuvers, the work in [15] focuses on anticipating traffic disruptions caused by incidents, whereas [16] proposes a cooperative algorithm designed to improve merging efficiency for vehicles entering highways from an on-ramp. These approaches demonstrate the benefits of cooperation in well-defined scenarios but are not intended to operate as general-purpose lane-change strategies under arbitrary traffic conditions.

In contrast to the above approaches, the proposed FORESEE model targets lane-change decision making as a continuous, decentralized, and traffic-oriented process. FORESEE leverages look-ahead V2X information to anticipate traffic evolution beyond immediate neighboring vehicles and evaluates lane-change decisions based on their impact on overall traffic efficiency rather than solely on local or short-term effects. By organizing lane usage according to anticipated traffic conditions while maintaining fully decentralized operation, FORESEE addresses an aspect that remains largely unexplored in existing strategy-level decentralized cooperative lane-change models.

Cooperative lane-change strategies can be further complemented by maneuver coordination mechanisms [17], particularly in situations where desired lane changes are hindered by insufficient gaps or conflicting vehicle intentions. Coordination enables vehicles to exchange information through V2X communications to facilitate mutually beneficial maneuvers. However, the effectiveness of coordinated maneuvers strongly depends on the underlying lane-change decision model. When coordination is combined with non-cooperative or purely local decision rules, maneuver outcomes may remain misaligned with global traffic objectives. Strategy-level cooperative models that explicitly consider traffic evolution, such as FORESEE, provide a suitable foundation for future research on coordinated V2X-enabled maneuvers, enabling tighter integration between decision-making and cooperative execution mechanisms.

## III. Traffic Analysis with Non-Cooperative Lane Change

We first analyze the traffic dynamics and performance using a non-cooperative strategy-level lane-change model. To this end, we employ MOBIL, a widely adopted decentralized lane-change model [9], which represents the reference approach for strategy-level decision-making under general traffic conditions. Using MOBIL as baseline allows us to isolate the impact of introducing cooperative look-ahead information and system-oriented decision criteria, while preserving the same decentralized and general-purpose modeling assumptions. The conducted analysis aims to explore how traffic behaves when vehicles do not leverage V2X data for their lane change decisions, and to identify scenarios where making use of V2X data could improve lane change decisions and enhance overall traffic flow.

### A. MOBIL

MOBIL uses two criteria to determine whether a lane change should be executed: the incentive criterion and the safety criterion. These criteria take into account the context of the ego vehicle $V_{ego}$ considering a lane change, the vehicle $V_{bo}$ behind $V_{ego}$ in its current lane, and the vehicle $V_{bn}$ behind $V_{ego}$ in the target lane. The incentive criterion is defined as follows:

$$(\tilde{a}_{ego} - a_{ego}) + p(\tilde{a}_{bo} - a_{bo} + \tilde{a}_{bn} - a_{bn}) > \Delta a_{th} \quad (1)$$

where $\tilde{a}_{ego}$, $\tilde{a}_{bo}$ and $\tilde{a}_{bn}$ denote the accelerations of $V_{ego}$, $V_{bo}$, and $V_{bn}$, respectively, if the lane change is executed. Conversely, $a_{ego}$, $a_{bo}$ and $a_{bn}$ represent their accelerations if the lane change is not executed. The parameter $p$, known as as the politeness factor, ranges from 0 to 1. The threshold $\Delta a_{th}$ determines the minimum acceleration gain required to justify a lane change, and is typically set to 0.2 m/s² [1]. Reducing $\Delta a_{th}$ increases the number of lane changes.

The safety criterion is defined as:

$$\tilde{a}_{bn} \geq b_{safe} \quad (2)$$

where $b_{safe}$ denotes the maximum deceleration that can be safely applied, and thus is a negative acceleration value. To meet the safety criterion, the deceleration experienced by $V_{bn}$ must not exceed $b_{safe}$ . If both the incentive and safety criteria are satisfied, $V_{ego}$ will proceed with the lane change. In the MOBIL model, the incentive criterion is designed to optimize the overall acceleration of all vehicles involved in the lane change. Consequently, the incentive criterion typically avoids any vehicle involved in a lane change from undergoing significant braking. On the other hand, the safety criterion focuses on preventing abrupt decelerations that could lead to unsafe traffic conditions.

### B. Simulation Framework

We analyze the impact of non-cooperative and cooperative mobility using a software platform that integrates traffic and V2X communications simulation. This platform uses the ns-3 network simulator for V2X communications and the VANET Highway Mobility module [18] to simulate the vehicular mobility. This module implements the Intelligent Driver Model (IDM) for longitudinal vehicle control [2]. For lateral control, it uses the MOBIL model for non-cooperative mobility (relying solely on local information) and the FORESEE proposal for cooperative mobility (leveraging V2X data). In this section, we focus exclusively on non-cooperative mobility utilizing the MOBIL model for lane change control.

We consider a 5 km highway scenario with six lanes (three lanes in each direction). The scenario incorporates periodic boundary conditions so that vehicles reaching one edge of the highway re-enter the simulation at the opposite edge. Throughout the simulation, the number of vehicles remains constant, maintaining a consistent average traffic density with possible localized variations. The number of vehicles is calculated to simulate various traffic densities, ranging from 10 vehicles per kilometer per lane (representing light traffic characterized by free flow and resilience to disturbances) to 40 vehicles per kilometer per lane (representing very dense, slow-moving traffic, susceptible to instabilities when faced with minor disturbances). For all densities evaluated, 80% of the vehicles are passenger cars and 20% are trucks. The desired speed of each vehicle is randomly determined following a uniform distribution with a possible deviation of ±20% around the average desired speed. The average desired speed is set to 33.3 m/s (120 km/h) for cars and 22.2 m/s (80 km/h) for trucks. At the beginning of the simulation, all the vehicles are distributed along the highway with constant spacing between the vehicles, with the restriction that only cars are allowed in the fastest left lane. We evaluate the highway scenario with and without a stationary obstacle in the middle of the right lane for each driving direction (Fig. 1). To ensure statistical accuracy, we conduct 40 different simulation runs for each scenario and vehicle density. Each simulation includes a warm-up period to allow traffic patterns to stabilize, and data is collected for over 1500 seconds following this warm-up period. Tables I and II provide the parameters used in our simulations for configuring the IDM and MOBIL models, respectively. MOBIL is configured to promote fully altruistic behavior in vehicles (i.e. $p$=1), which is deemed to improve the traffic flow [1].

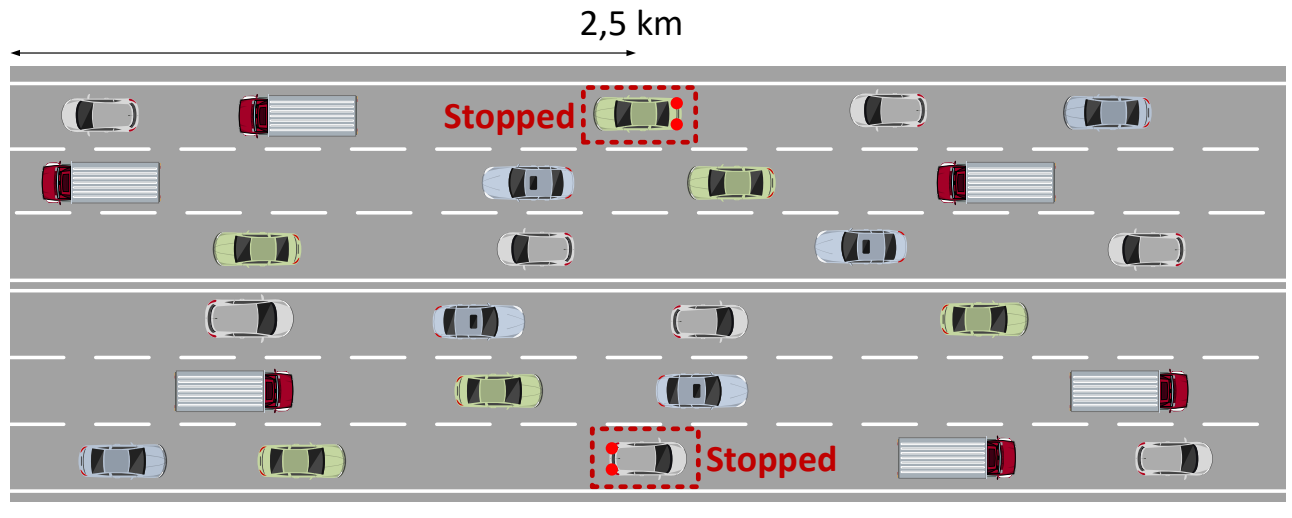

Fig. 1. Highway scenario with obstacle in the right lane.

TABLE I. IDM Parameters

| | Trucks | Cars |
|---|---|---|
| Desired speed ($v_0$) | 22.2 m/s ±20% | 33.3 m/s ±20% |
| Time headway ($T$) | 1 s | 0.8 s |
| Minimum gap ($s_0$) | 2 m | 2 m |
| Maximum acceleration ($a$) | 1.5 m/s² | 1.5 m/s² |
| Comfortable deceleration ($b$) | 2 m/s² | 2 m/s² |

TABLE II. PARAMETERS USED FOR MOBIL MODEL

| | |
|---|---|
| Lane change threshold ($\Delta a_{th}$) | 0.2 m/s² |
| Politeness factor ($p$) | 1 |
| Maximum safe deceleration ($b_{safe}$) | -4 m/s² |

## C. *Traffic analysis*

Fig. 2 illustrates the evolution over time of the average desired speed for vehicles in each lane when using the MOBIL model, i.e. with non-cooperative lane change decisions based solely on local driving information. The figure reveals a higher average speed in the left lane, as trucks are prohibited from using this lane in our simulations. The middle and right lanes, available to both cars and trucks, show no significant difference in the average desired speeds of vehicles. Fig. 3 illustrates the difference between the desired and actual speeds of vehicles across varying vehicle densities. In the box plot, the red line indicates the average of these differences. The boundaries of the box correspond to the 10th and 90th percentiles, while the whiskers extend to the 1st and 99th percentiles. The figure reveals that the discrepancy between desired and actual speeds is notably high, even at low vehicle densities. The figure shows that, up to a density of 15 vehicles per kilometer per lane, vehicles achieve their desired speeds only 10% of the time. This percentage decreases to just 1% at higher densities. Moreover, Fig. 3 shows that at densities of 15 vehicles per kilometer per lane or higher, the largest speed differences (90th and 99th percentile) exceed 40 km/h. Reducing the lane change threshold ($\Delta a_{th}$) to 0.03 m/s² increases the number of lane changes, as shown in Fig. 4. However, this increase in lane changes does not reduce the gap between desired and actual speeds. The trends and values observed with a lower $\Delta a_{th}$ remain consistent with those depicted in Fig. 3.

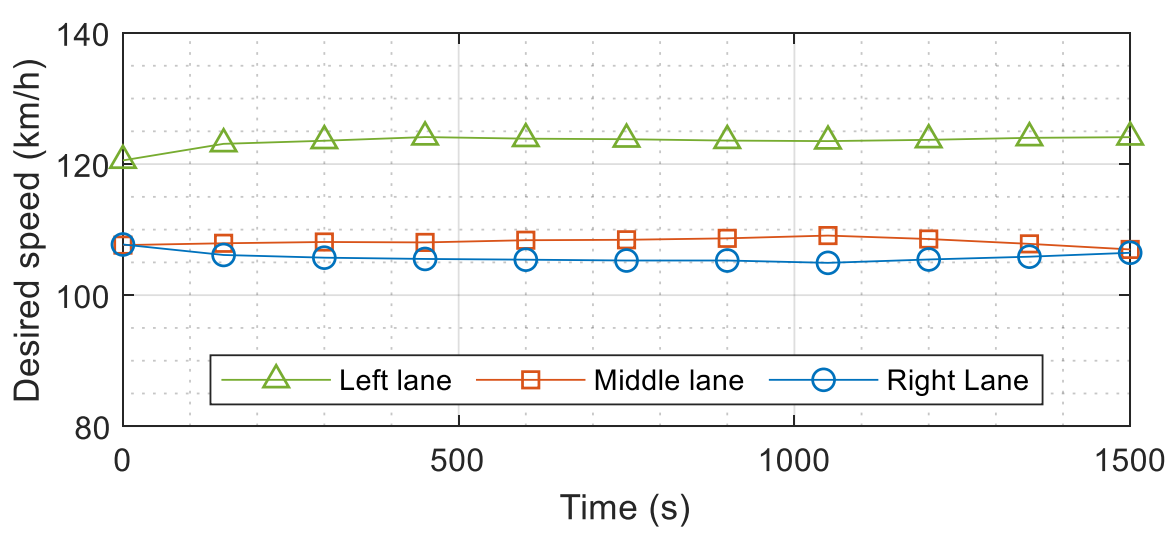

Fig. 2. Average desired speed of all vehicles in each lane as a function of the time. Scenario without obstacles with 20 vehicles/km/lane density.

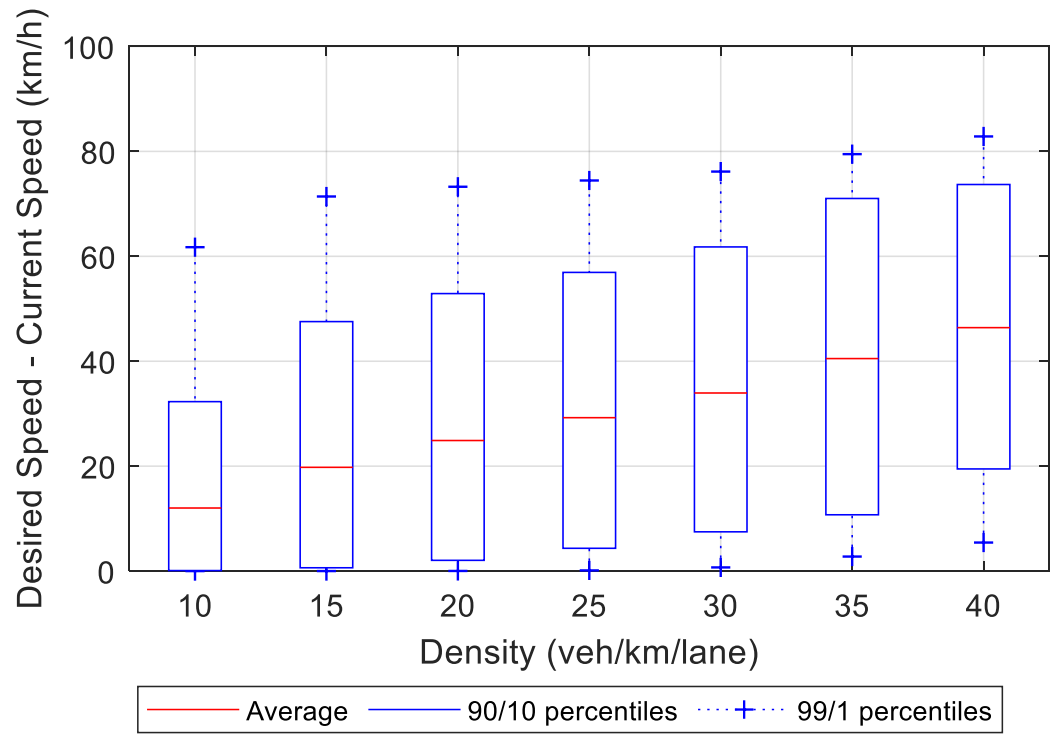

Fig. 3. Box plot of the difference between desired and actual speeds as a function of the density of vehicles. Scenario without obstacles.

This is because many of the additional lane changes triggered by reducing $\Delta a_{th}$ from 0.2 m/s² to 0.03 m/s² result in vehicles quickly reverting to their original lanes. Fig. 5 illustrates this behavior by showing the average number of lane changes per vehicle and hour that were followed by a return to the initial lane within 2, 5, and 10 seconds. Fig. 4 shows that reducing $\Delta a_{th}$ from 0.2 m/s² to 0.03 m/s² increases the number of lane changes. However, many of these additional lane changes are ineffective, as vehicles often return to their original lane shortly after the maneuver (Fig. 15). This suggests that merely increasing the number of lane changes does not inherently enhance traffic flow unless those changes are well-configured and strategically executed. Moreover, Fig. 6 shows that decreasing the lane change threshold does not mitigate the impact of road obstacles either. The figure shows the number of vehicles stuck behind the road obstacle introduced in the second traffic scenario. As vehicle density increases, the number of vehicles unable to bypass the obstacle increases significantly, due to fewer available opportunities to change lanes. In non-cooperative mobility models, vehicles rely solely on local information to make lane change decisions and cannot anticipate traffic conditions further ahead. Vehicles start seeking lane changes only upon experiencing the obstruction's impact, i.e. when encountering the road obstacle or the congestion that is being formed. At this point, it is typically too late to find suitable gaps for effective lane changes, leading to more vehicles getting stuck behind the obstacle.

The analysis in this section highlights the need and opportunity for cooperative driving as it can exploit look-ahead information from V2X communications for better driving decisions. Developing and testing such cooperative driving solutions necessitates mobility models capable of effectively utilizing this V2X data. To bridge this gap, we propose FORESEE, a cooperative lane change model that leverages V2X data for lane change decisions.

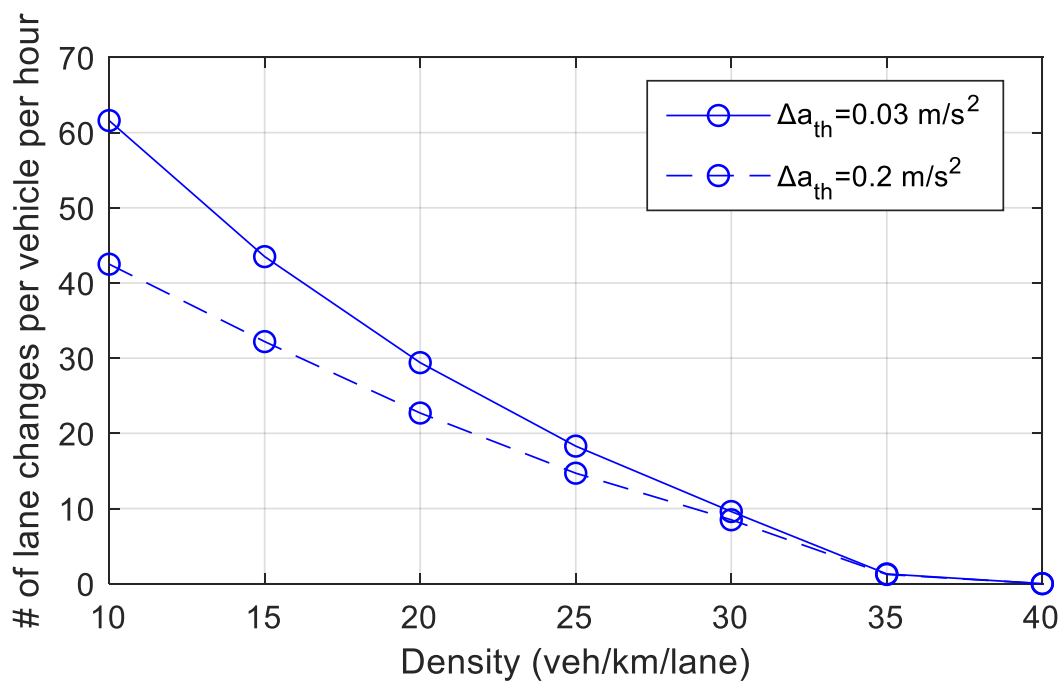

Fig. 4. Impact of $\Delta a_{th}$ on the average number of lane changes normalized per vehicle and per hour of simulation time. Scenario without obstacles.

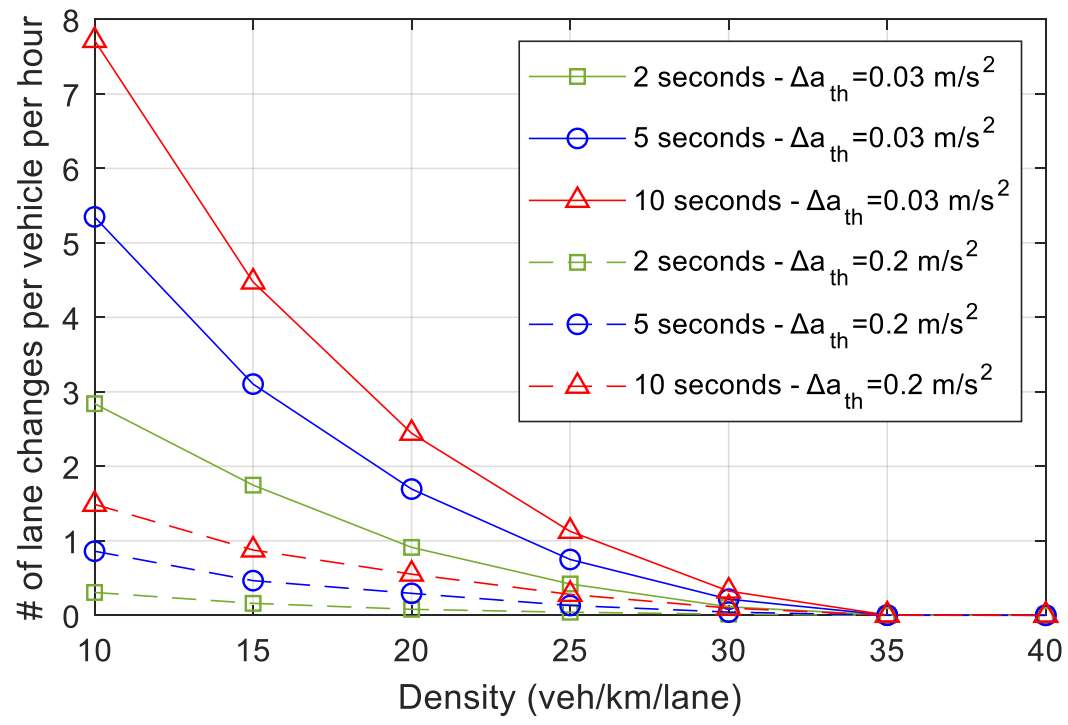

Fig. 5. Average number of lane changes per vehicle and hour that were followed by a return to the initial lane within 2, 5, and 10 seconds. Scenario without obstacles

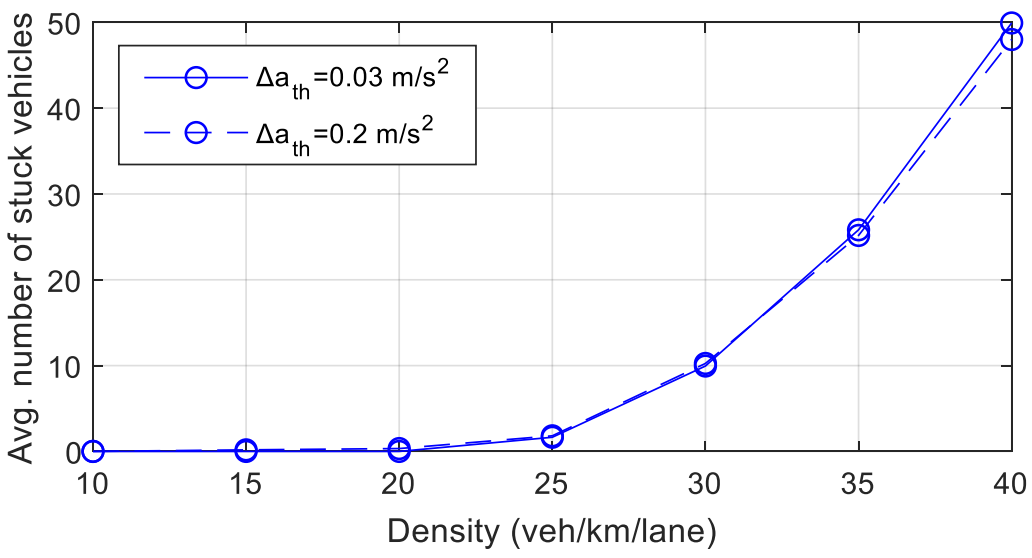

Fig. 6. Number of vehicles stuck behind the road obstacle. Scenario with obstacles.

## IV. Cooperative Look-Ahead Lane Change Model

The general aim of traffic optimization is to achieve homogenization of traffic flow, which is considered the Golden Rule of Traffic Flow Optimization [1]. Homogenized traffic helps prevent local and temporary reductions in road capacity and address perturbations in traffic flow. Such disruptions are often caused by abrupt lane changes or braking maneuvers, which can be triggered by differences in speed between vehicles traveling in the same lane. In this context, we propose FORESEE (look-ahead lane change for speed-based vehicle ordering), a lane change model designed to reduce speed differences by organizing vehicles into lanes based on their desired speeds. The objective of FORESEE is to foster traffic homogenization by directing vehicles with lower desired speeds to the right lanes and those with higher desired speeds towards the left lanes (Fig. 7). To this aim, FORESEE leverages V2X data to estimate the speed of each lane and make strategic lane change decisions based on anticipated traffic conditions. By using this look-ahead data, FORESEE can effectively organize lane changes that enhance overall traffic flow and reduce the likelihood of bottlenecks.

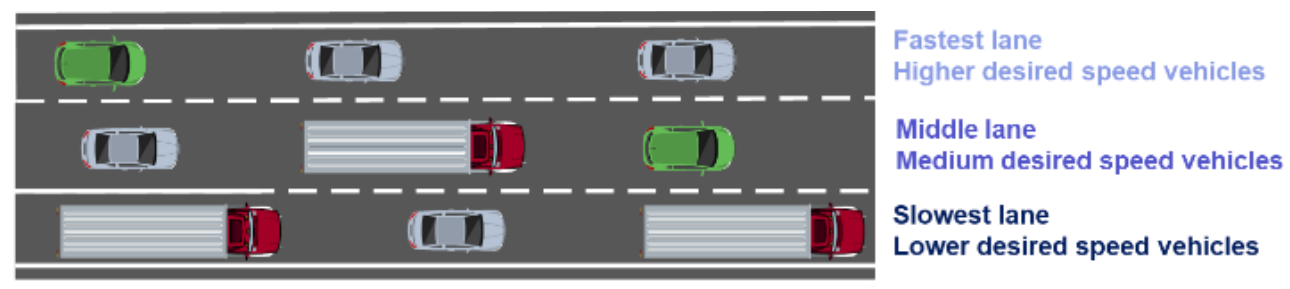

Fig. 7. Traffic ordering based on desired speed with FORESEE.

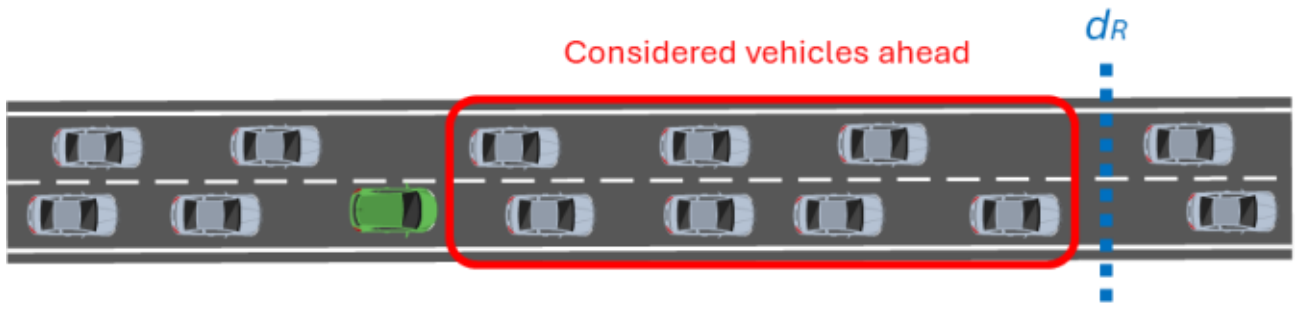

Fig. 8. V2X look-ahead information used in FORESEE.

### A. FORESEE Lane Change Model

FORESEE uses information received from surrounding vehicles via V2X communications to make its lane change decisions. Specifically, it utilizes the position and speed of all neighboring vehicles within a defined range $d_R$ from the ego vehicle (Fig. 8), where positional information is mainly used to identify the lane of each vehicle and therefore only requires lane-level accuracy. This information can be obtained, for example, from Cooperative Awareness Messages (CAMs) or Basic Safety Messages (BSMs), which are continuously broadcasted by cooperative, connected and automated vehicles (CAVs). Using V2X information from nearby vehicles, FORESEE can anticipate traffic conditions and make more strategic lane change decisions based on a better understanding of the surrounding traffic. This is in contrast to non-cooperative lane change models, which rely solely on local information, e.g. the vehicles directly behind or in front of the ego vehicle. In particular, FORESEE estimates the speed of lanes to evaluate the surrounding traffic conditions, and decide about a possible lane change. FORESEE estimates the speed of a lane $v_{lane}$ as the speed of the slowest vehicle in the lane within a range ahead $d_R$ (Fig. 8):

$$v_{lane} = \min_i(v_i) \tag{3}$$

where $v_i$ represents the speed of vehicle $V_i$ in the lane under evaluation. $v_i$ is obtained through the V2X messages broadcasted by nearby vehicles. By considering the minimum speed within the range $d_R$, $v_{lane}$ represents an estimate of the speed that vehicles in the lane will likely experience in the near future, as they will need to adjust to the speed of the slowest vehicle ahead. Using FORESEE, an ego vehicle $V_{ego}$ estimates the speed of its current lane ($v_{lane}$), the speed of its adjacent left lane ($v_{left}$), and the speed of its adjacent right lane ($v_{right}$). It then applies an incentive criterion and a comfort criterion to assess whether a lane change is desirable and feasible, respectively. A lane change is executed only when both criteria are satisfied. In this regard, the structure of FORESEE's decision process is fully aligned with the established methodological framework used in state-of-the-art strategy-level models like MOBIL.

The incentive criterion, depicted in the flowchart in Fig. 9, determines whether a lane change aligns with FORESEE's goal of directing slower vehicles to the right lanes and faster vehicles to the left lanes. An ego vehicle first evaluates if it should consider a change to the right lane by comparing $v_{lane}$ and $v_{right}$:

$$|v_{right} - v_{lane}| > \delta_{ls} \tag{4}$$

where $\delta_{ls}$ is a threshold to prevent frequent lane changes within short time intervals between adjacent lanes with similar speeds. The ego vehicle will not consider a lane change if the estimated speeds of its current lane and the adjacent right lane are too similar. If it is not the case, a change to the right lane becomes desirable if the right lane is faster than the current lane. This is because of FORESEE's goal of directing slower vehicles to the right lanes. If the adjacent right lane is slower than the current lane, $V_{ego}$ will only change to the right lane if its desired speed $v_0$ is sufficiently low, indicating it should be in the slower right lane. This condition is expressed as follows:

$$v_0 < DSth_{right} - \delta_{ds} \tag{5}$$

where $v_0$ is the desired speed of $V_{ego}$, $DSth_{right}$ is the speed threshold required to initiate a lane change to the right, and $\delta_{ds}$ is a threshold aimed at preventing frequent lane changes between the two lanes if $v_0$ and $DSth_{right}$ are similar. $DSth_{right}$ is defined as:

$$DSth_{right} = v_{right} \cdot (1 + \rho) \tag{6}$$

where $\rho$ represents an offset with respect to $v_{right}$, the estimated speed of the right lane. The parameter $\rho$ opens the door for vehicles to move to the right lane even if their desired speed is higher than the speed of that lane. This is necessary because otherwise, only a few vehicles would satisfy the incentive conditions to move to the right lanes, which could imbalance lane occupancy and increase the risk of congestion in the left lanes.

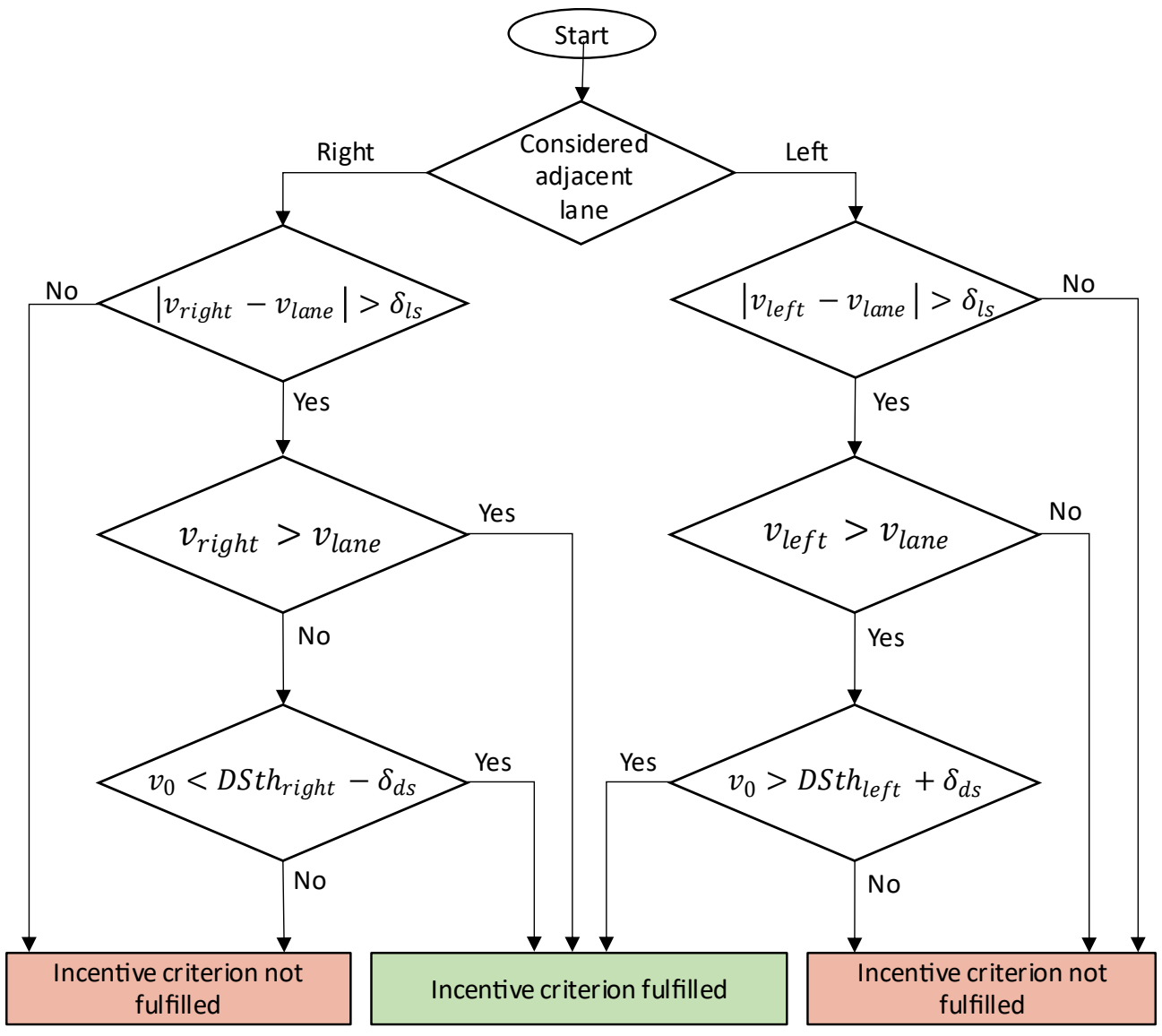

Fig. 9. Incentive criterion in FORESEE.

The opposite conditions apply when the ego vehicle $V_{ego}$ considers changing to the adjacent left lane (see Fig. 9), directing the fastest vehicles towards the left lanes. In this case, eq. (4), (5) and (6) are modified as follows:

$$|v_{left} - v_{lane}| > \delta_{ls} \tag{7}$$

$$v_0 > DSth_{left} + \delta_{ds} \tag{8}$$

$$DSth_{left} = v_{lane} \cdot (1 + \rho) \tag{9}$$

where $v_{left}$ is the estimated speed of the left lane and $DSth_{lane}$ is the speed threshold to move to the left lane. We utilize the same thresholds $\delta_{ls}$ and $\delta_{ds}$ for the right and left lane changes since these thresholds have been defined to avoid unstable vehicle behavior characterized by frequent lane changes within short time intervals between two adjacent lanes with similar speeds.

The comfort criterion is designed to ensure that lane changes do not introduce significant traffic instabilities. We define $b_{comfort}$ as the minimum (negative) acceleration that vehicles involved in a lane change can handle without causing discomfort or instability. With FORESEE, the vehicles that may need to decelerate following a lane change are the ego vehicle $V_{ego}$ and the vehicle $V_{bn}$ immediately behind $V_{ego}$ in the adjacent (new) lane. The comfort criterion is satisfied if the following condition holds:

$$\tilde{a}_{ego} \ , \ \tilde{a}_{bn} \geq b_{comfort} \tag{10}$$

where $\tilde{a}_{ego}$ and $\tilde{a}_{bn}$ represent the accelerations that $V_{ego}$ and $V_{bn}$ would experience after the lane change under consideration. The comfort criterion ensures a safe lane change for $V_{ego}$ and $V_{bn}$ in the adjacent lane, and it prevents harsh braking for the involved vehicles even if it does not pose a safety risk, as it could disrupt traffic flow and impair traffic efficiency.

### B. *Illustration of non-cooperative (or local) and cooperative lane changes*

This section illustrates the impact of non-cooperative and cooperative lane changes on vehicle behavior using the MOBIL and FORESEE lane change models in the following three representative scenarios:

- Scenario 1 (Fig. 10.a): the ego vehicle $V_{ego}$ is a slow truck traveling in the middle lane.
- Scenario 2 (Fig. 11.a): $V_{ego}$ is a fast vehicle traveling in the right lane with some empty space ahead.
- Scenario 3 (Fig. 12.a): $V_{ego}$ is a fast vehicle traveling in the left lane, while the middle lane contains slower vehicles but offers more empty space ahead.

Note that in Fig. 10, Fig. 11, and Fig. 12 all vehicles within range are visible and marked with a wireless connectivity symbol; FORESEE uses information from all connected vehicles in range, whereas MOBIL considers only the adjacent vehicles $V_{bn}$, $V_{fn}$, $V_{bo}$, and $V_{fo}$. Fig. 10.a illustrates the first scenario where $V_{ego}$ is a slow truck traveling in the middle lane. With MOBIL (Fig. 10.b), $V_{ego}$ does not change to the right lane despite having enough space between $V_{bn}$ and $V_{fn}$, as $V_{ego}$ does not have an incentive to do so. This is because neither $V_{ego}$ nor $V_{bo}$ would increase their acceleration with the lane change since they both have some free space ahead in their current lane. Fig. 10.c depicts the behavior of vehicles with FORESEE. In this case, $V_{ego}$ receives speed information from nearby vehicles via V2X communication and detects that it is slower than $V_{fo}$ and its desired speed is not high enough to justify remaining in the middle lane. $V_{ego}$ then switches to the right lane as soon it finds sufficient gap (ahead of $V_{bn}$).

Fig. 11.a depicts the second scenario, where $V_{ego}$ is a fast vehicle traveling in the right lane with a significant gap ahead. With MOBIL (Fig. 11.b), $V_{ego}$ opts to remain in the right lane despite having a sufficient gap to change to the middle lane, as the gap ahead in the right lane is larger. Initially, $V_{ego}$ increases its acceleration and speed compared to the vehicles in the middle lane. However, it would later need to decelerate as $V_{fo}$ is a slower vehicle. Changing to the middle lane would have avoided this later deceleration and offered $V_{ego}$ the opportunity to change to the leftmost lane, where trucks are not allowed. With FORESEE (Fig. 11.c), $V_{ego}$ detects that $V_{fo}$ is a slower vehicle thanks to the speed information received via V2X communications, and transitions to the middle lane as it better aligns with its desired speed. In this case, both $V_{ego}$ and $V_{bn}$ would need to decelerate instantaneously due to the narrower gap between $V_{bn}$ and $V_{fn}$ compared to the gap in the right lane. However, $V_{ego}$ avoids the later deceleration as it anticipates traffic conditions ahead and detects that $V_{fo}$ is a slower vehicle. Additionally, $V_{ego}$ can change to the left lane if its desired speed aligns better with the speed in the left lane than in the middle lane.

Fig. 12.a illustrates the third scenario, where $V_{ego}$ is a fast vehicle in the left lane, while there is a substantial gap ahead in the middle lane. With MOBIL (Fig. 12.b), $V_{ego}$ opts to change to the middle lane to increase its acceleration by utilizing the available gap. This lane change also allows $V_{ego}$ to leave more space for $V_{bo}$ to accelerate. However, since $V_{fn}$ is a slower vehicle, $V_{ego}$ will need to decelerate as it approaches $V_{fn}$ or consider changing lanes again. With FORESEE (Fig. 12.c), $V_{ego}$ opts to remain in the left lane, despite a larger gap ahead in the middle lane where it could accelerate instantly. This is because $V_{ego}$ detects that $V_{fn}$ is a slower vehicle thanks to the speed information received via V2X communications, and anticipates that it would need to decelerate later if it changes lanes. So, looking ahead, $V_{ego}$ identifies that its desired speed aligns better with the conditions in the left lane.

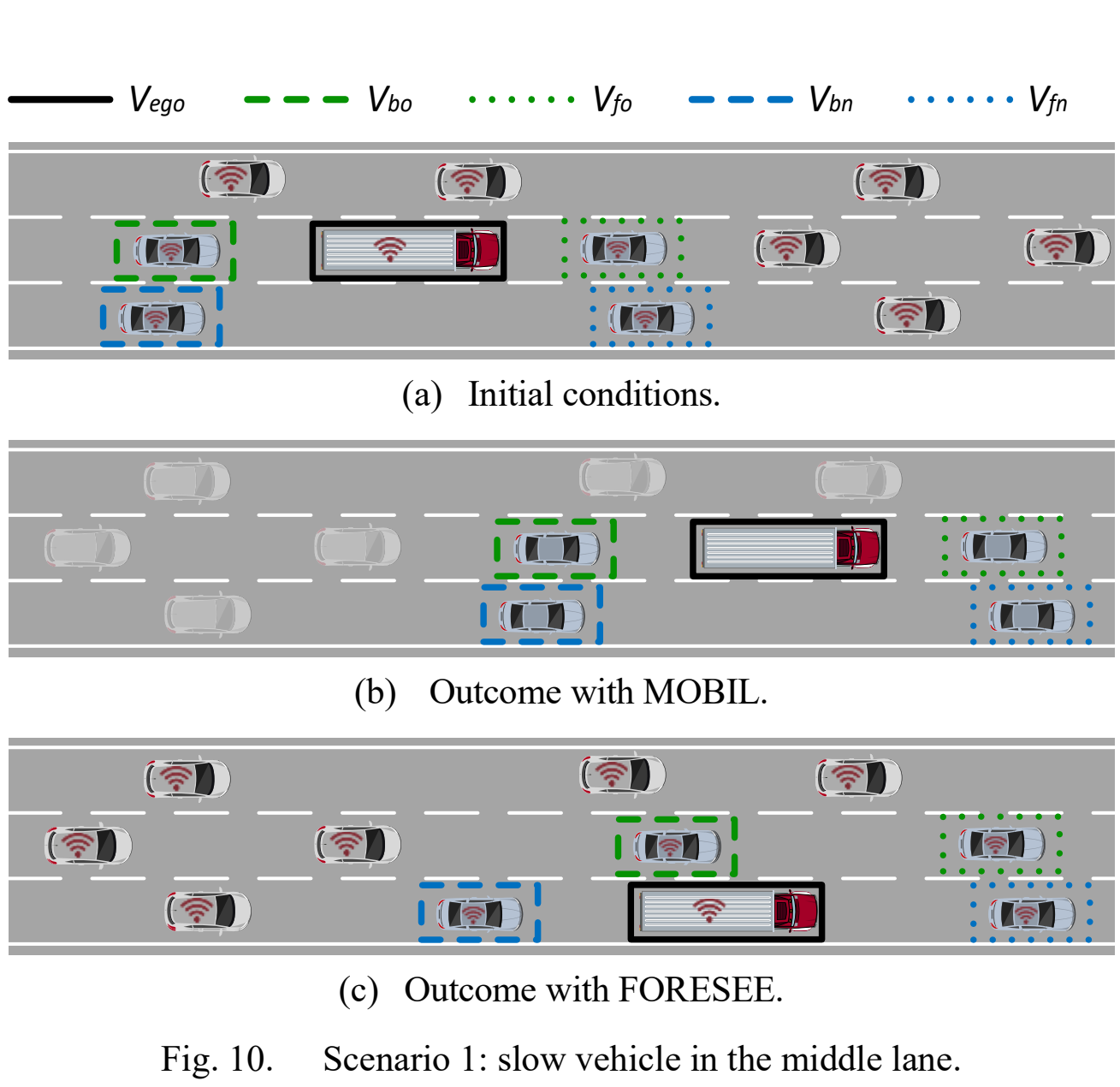

(a) Initial conditions.

(b) Outcome with MOBIL.

(c) Outcome with FORESEE.

Fig. 10. Scenario 1: slow vehicle in the middle lane.

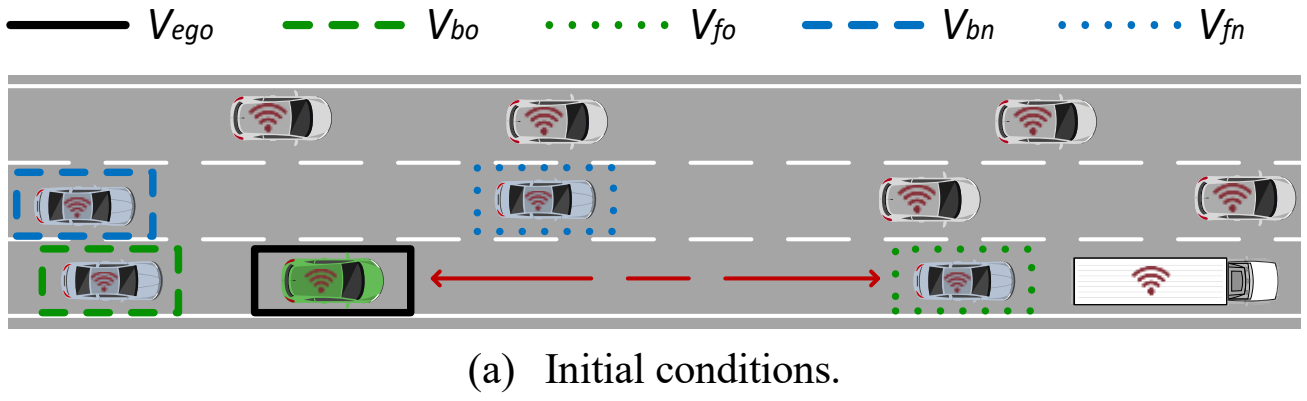

(a) Initial conditions.

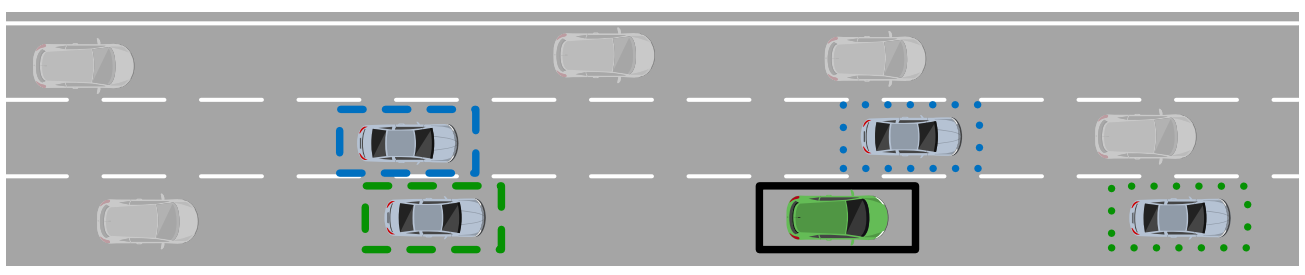

(b) Outcome with MOBIL.

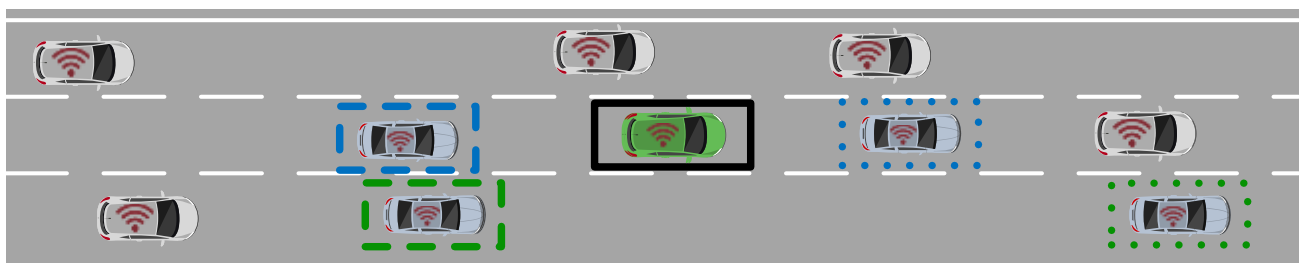

(c) Outcome with FORESEE.

Fig. 11. Scenario 2: fast vehicle in the right lane with empty space ahead.

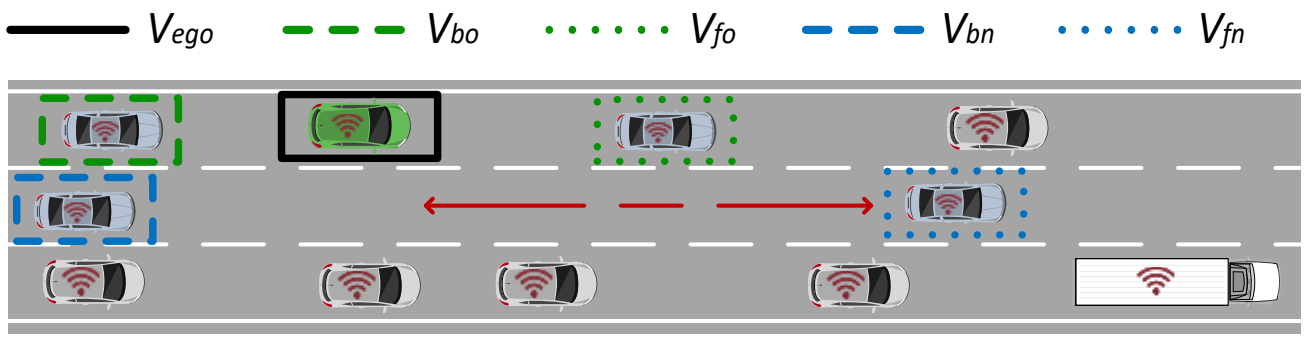

(a) Initial conditions.

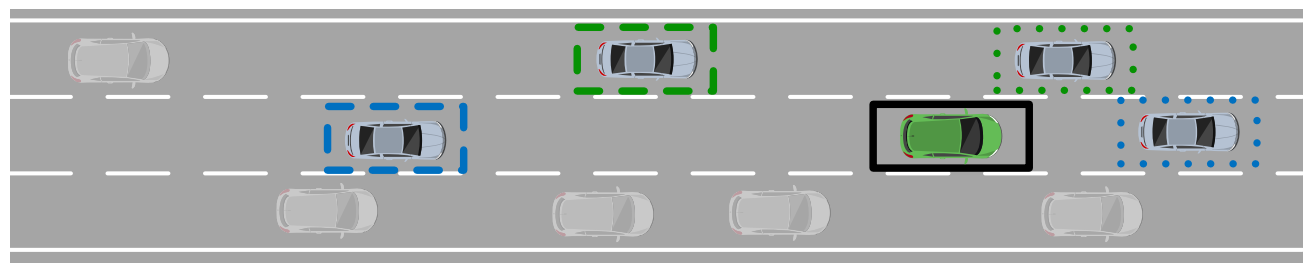

(b) Outcome with MOBIL.

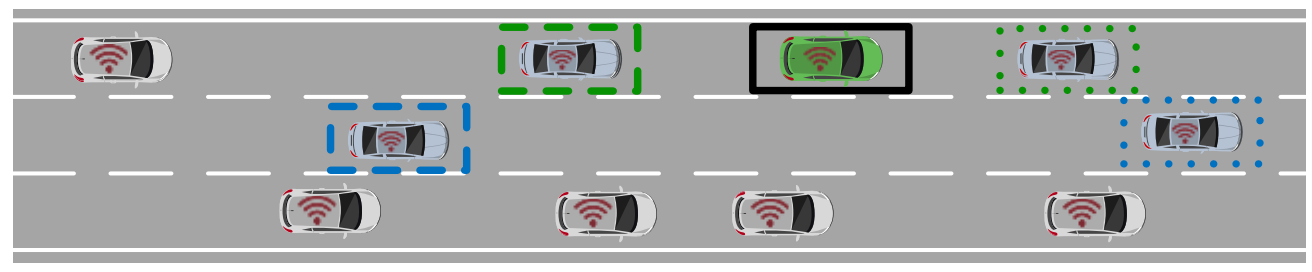

(c) Outcome with FORESEE.

Fig. 12. Scenario 3: fast vehicle in the left lane.

## V. IMPACT OF COOPERATIVE LANE CHANGES ON TRAFFIC

This section analyzes the impact of cooperative lane changes on traffic using the proposed FORESEE model. Table III details the values of the model's parameters used in our evaluation. These parameters have been carefully configured to ensure optimal performance across all studied scenarios and traffic densities. The evaluation is done considering the simulation framework presented in Section III.B.

TABLE III. FORESEE PARAMETERS

| | |
|---|---|
| Range ($d_R$) | 500 m |
| Offset ($\rho$) | 0.3 |
| Maximum comfort deceleration ($b_{comfort}$) | -3 m/s² |
| Lane speed difference margin ($\delta_{ls}$) | 0.5 m/s |
| Desired speed threshold margin ($\delta_{ds}$) | 0.5 m/s |

Although the FORESEE range $d_R$ is set to 500 m, the model does not assume perfect communication up to that distance. Packet losses are fully accounted for through the ns-3 V2X simulation integrated into our framework, and the effective information available to each vehicle reflects the actual PDR and update rate resulting from the communication model. Different ranges have been tested, but the selected one offers a good balance between communication reliability and range. Additional simulations with partial V2X penetration (50%) were conducted, confirming that the proposed cooperative strategy continues to provide benefits and preserves the observed behavioral trends under mixed traffic conditions

### A. Traffic dynamics

Fig. 13 illustrates how the average desired speed for vehicles in each lane evolves over time using the FORESEE model. The figure includes results for full V2X penetration as well as partial V2X penetration (50% penetration rate) where only half of the vehicles use V2X communication and the FORESEE lane change strategy. The other half of the vehicles use MOBIL and do not have V2X communications. FORESEE leverages V2X data to estimate lane speeds and direct connected vehicles to lanes based on their desired speeds. The model leverages V2X data to estimate lane speeds and direct vehicles to lanes based on their desired speeds. As depicted in Fig. 13, FORESEE efficiently organizes vehicles into lanes according to their speed preferences: those with higher desired speeds are routed to the left lane, while those with lower speeds are placed in the right lane. This organization enhances the desired speed in the left lane compared to scenarios where lane changes are based solely on local information (Fig. 2). In the latter case, vehicles had similar speeds in the right and middle lanes, suggesting that slow vehicles move interchangeably between these lanes, which in turn affects the speed of the faster lane. Under partial penetration, this ordering is still preserved, although with a reduced magnitude due to the presence of non-connected vehicles relying solely on local information. Since this lane-level ordering constitutes the fundamental mechanism driving the performance improvements analyzed in the remainder of the paper, additional partial-penetration results are not repeated for the other metrics in order to preserve clarity and focus.

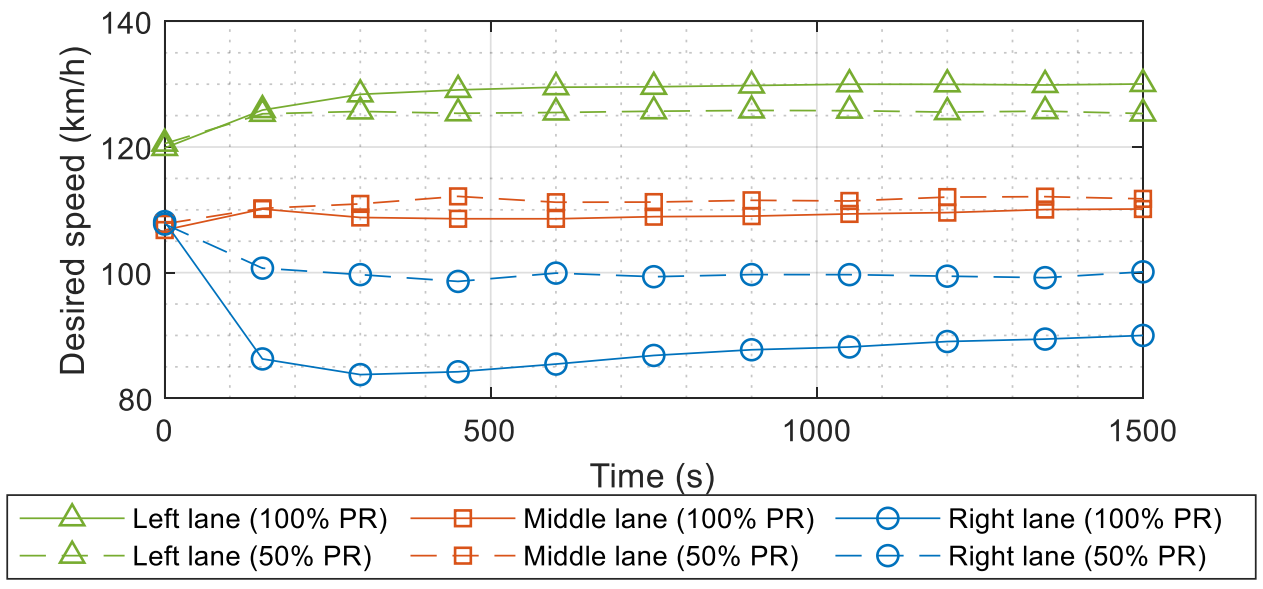

Fig. 13. Average desired speed of all vehicles in each lane as a function of the simulation time obtained with FORESEE with 20 vehicles/km/lane density. Results for full V2X penetration rate (100% PR) and partial penetration rate (50% PR)

Fig. 14 compares the difference between the desired and actual speeds of vehicles across varying vehicle densities using non-cooperative (MOBIL) and cooperative (FORESEE) lane changes. The figure demonstrates that cooperative lane changes, leveraging look-ahead V2X data to anticipate traffic conditions and organize lane changes, allow vehicles to maintain speeds closer to their desired levels. This effect is most pronounced at low and medium densities, where lane changes are more feasible, and diminishes at very high densities. In high congestion scenarios, limited opportunities for lane changes lead to inevitable larger differences between desired and actual speeds. Additionally, Fig. 14 shows that cooperative lane changes significantly reduce the variability in the difference between desired and actual speeds, which in turn reduces the variability in actual speeds. This reduces accelerations and decelerations, enhancing both driving comfort and efficiency. This is evident in Fig. 15, which shows that cooperative lane changes decrease the absolute value of vehicles' accelerations compared to non-cooperative lane changes.

We also analyze the traffic dynamics of cooperative lane changes in the presence of road obstacles. Fig. 16 compares the average number of vehicles stuck behind a road obstacle when using cooperative (FORESEE) and non-cooperative (MOBIL) lane changes. When an obstacle appears in the right lane, vehicles attempt to merge into the middle lane to bypass it. In non-cooperative scenarios, vehicles do not attempt to change lanes until they are directly affected by the traffic disturbance, resulting in limited opportunities to maneuver.

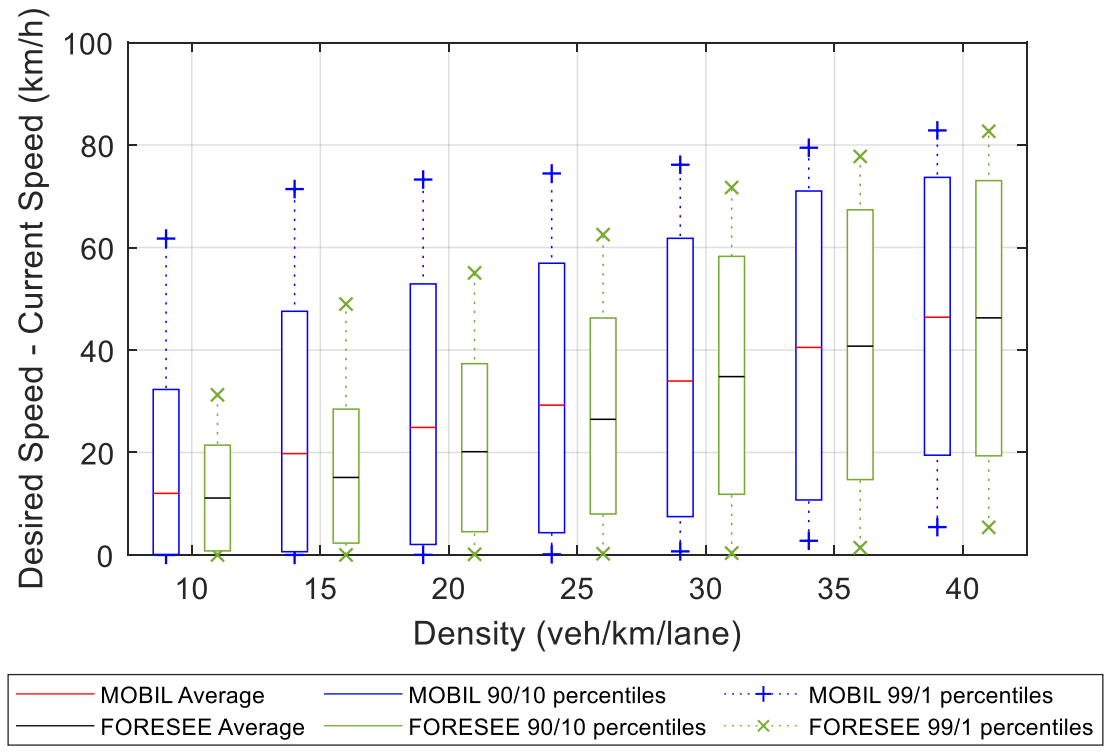

Fig. 14. Box plots of the difference between desired and actual speeds as a function of the density of vehicles with non-cooperative (MOBIL) and cooperative (FORESEE) lane changes.

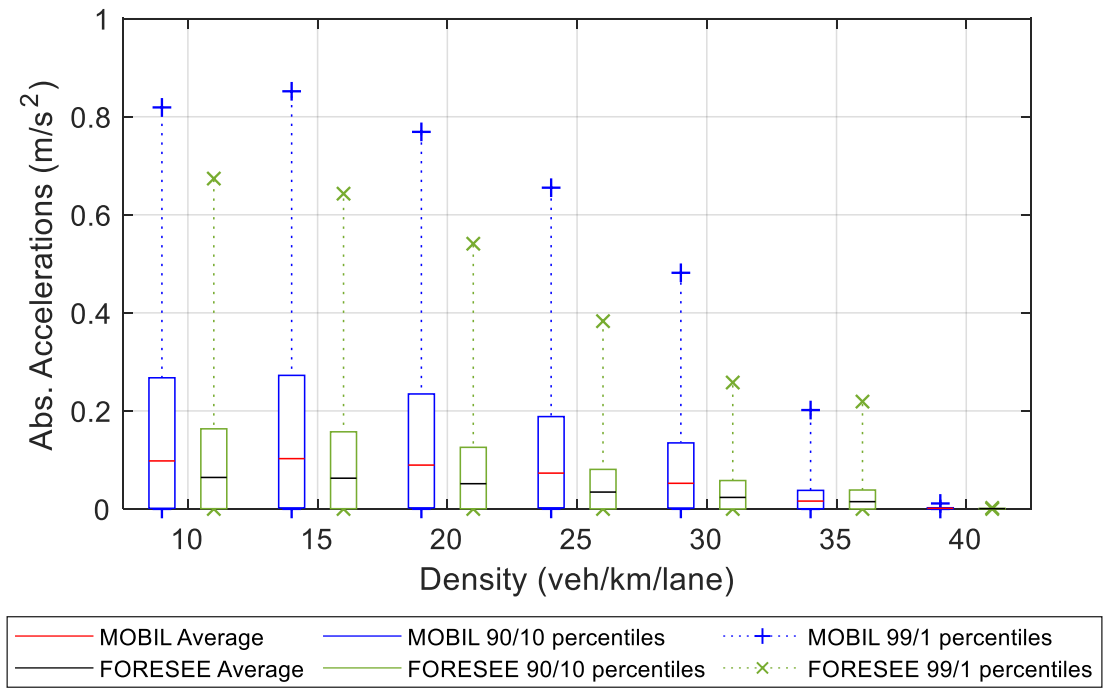

Fig. 15. Box plots of the vehicle's accelerations (absolute value) as a function of the density of vehicles with non-cooperative (MOBIL) and cooperative (FORESEE) lane changes.

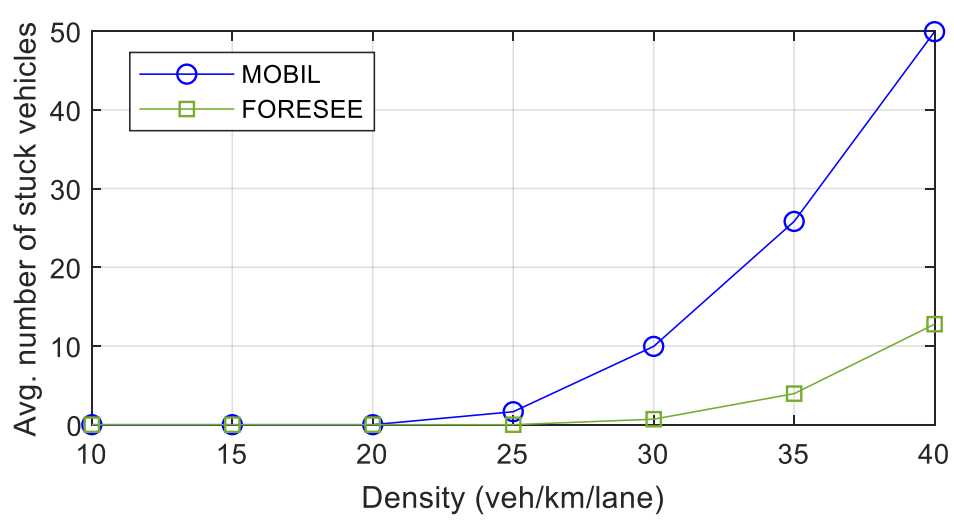

Fig. 16. Number of vehicles stuck behind the road obstacle as a function of the vehicle density.

On the other hand, vehicles utilizing cooperative lane changes leverage V2X data to anticipate obstacles and organize lane changes to minimize disruption. This is illustrated in Fig. 17, which shows the average distance from the obstacle at which vehicles switch from the right to the middle lane. For non-cooperative lane changes, this distance increases with traffic density, as decisions are based solely on local information. At low densities, vehicles change lanes only when close to the obstacle because its traffic impact is minimal and only noticeable at short distances. As density increases, the traffic impact grows, leading to more vehicles getting stuck behind the obstacle (Fig. 16). In such cases, non-cooperative vehicles change lanes at greater distances, as the long queue formed behind the obstacle affects them sooner. In contrast, Fig. 17 reveals that cooperative vehicles anticipate obstacles well in advance, regardless of traffic density, allowing them to manage lane changes efficiently and significantly reduce the number of vehicles stuck behind the obstacle (by a factor of 4 to 6). Additionally, Fig. 16 demonstrates that cooperative lane changes can effectively mitigate the impact of obstacles even at higher vehicle densities compared to non-cooperative lane changes.

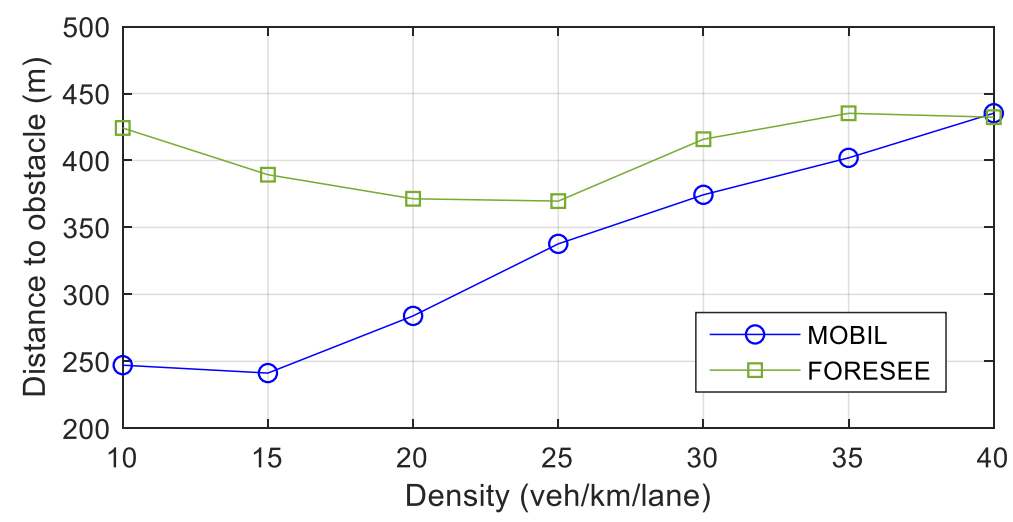

Fig. 17. Distance to the obstacle at which approaching vehicles change from the right to the middle lane as a function of the vehicle density.

## *B. Traffic efficiency*

Fig. 18 illustrates the impact of cooperative and non-cooperative lane changes on average vehicle speeds across different traffic densities. Fig. 18.a compares the overall average speed for all vehicles in the scenario. Fig. 18.b and Fig. 18.c depict box plots of the average speeds of all cars and trucks, respectively. The limits of the box represent the 10th and 90th percentiles, while the whiskers represent the 1st and 99th percentiles.

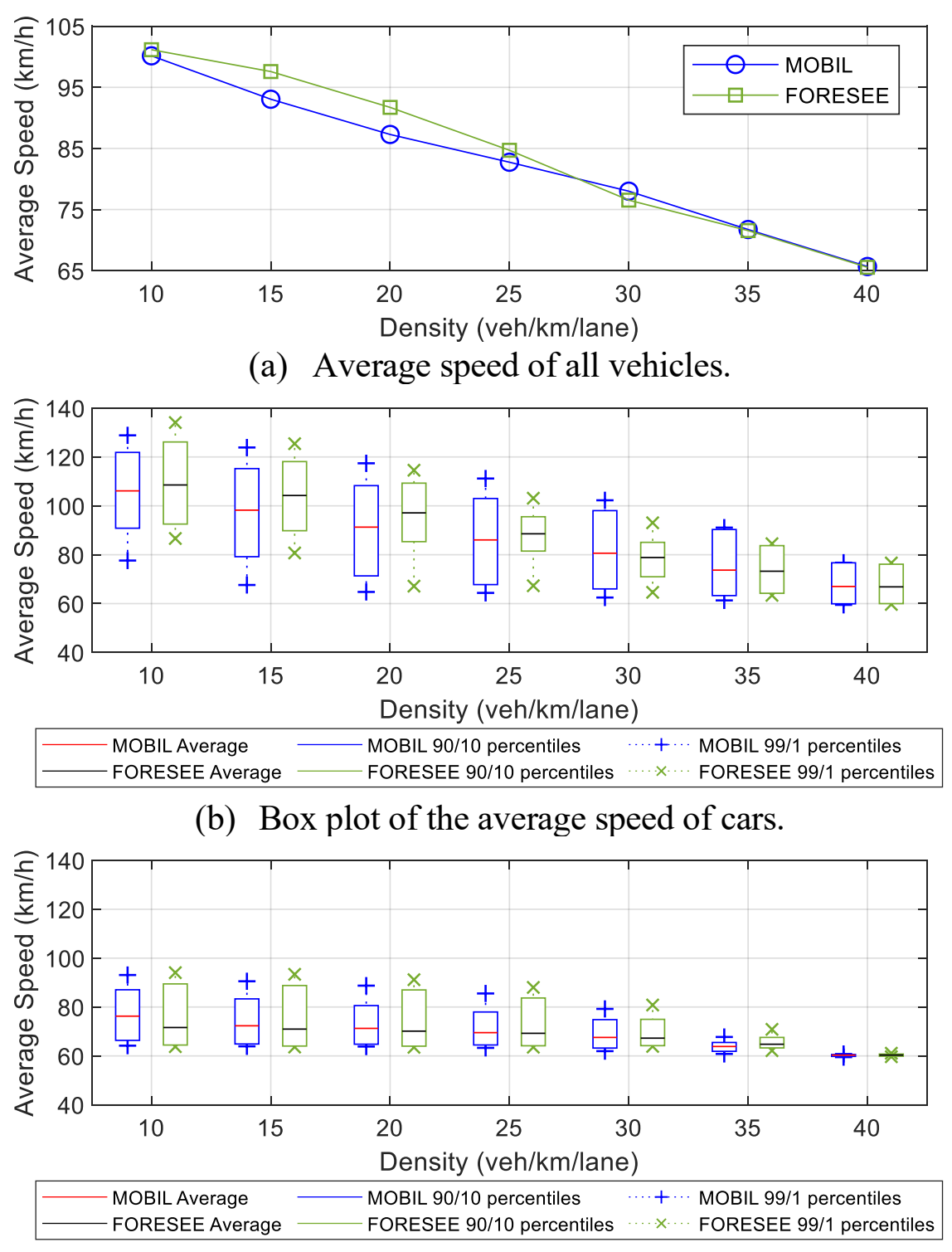

(a) Average speed of all vehicles.

(b) Box plot of the average speed of cars.

(c) Box plot of the average speed of trucks.

Fig. 18. Average speed with cooperative (FORESEE) and non-cooperative (MOBIL) lane changes as a function of the density.

Fig. 18.a demonstrates that the ability to strategically organize lane changes using look-ahead V2X data increases the average speeds of vehicles. For instance, at a density of 20 vehicles per kilometer per lane, the average speed is 91.7 km/h with cooperative lane changes (FORESEE) compared to 87.2 km/h with non-cooperative lane changes (MOBIL). This improvement in average speeds with cooperative lane changes translates to an increase in road capacity by 532 vehicles per hour. It is important to note that this improvement in average speeds is due to a more efficient distribution of vehicles across lanes rather than a higher number of lane changes. This is evidenced in Fig. 19, which shows that FORESEE results in fewer lane changes compared to MOBIL. Additionally, Fig. 18.a shows that there are no significant differences in average speeds between the two lane change models at very low or very high traffic densities. In low-density scenarios, vehicles can move freely between lanes and maintain speeds close to their desired levels. As a result, the method used to decide on lane changes does not impact the average speed. In high-density conditions, the limited availability of gaps restricts the opportunities for lane changes, resulting in similar average speeds for both models. Fig. 18.b shows that cooperative lane changes reduce the variability of the speed of cars as FORESEE uses V2X communications to anticipate road traffic conditions and organize vehicles into lanes based on their desired speeds. For trucks, this approach results in most trucks being routed to the right lane (Fig. 18.c), leading to a slight decrease in their average speed. However, approximately 20% of trucks with higher desired speeds can still access the middle lane with FORESEE. This explains the slightly larger variability in truck speeds compared to non-cooperative lane changes, where trucks move interchangeably between the right and middle lanes.

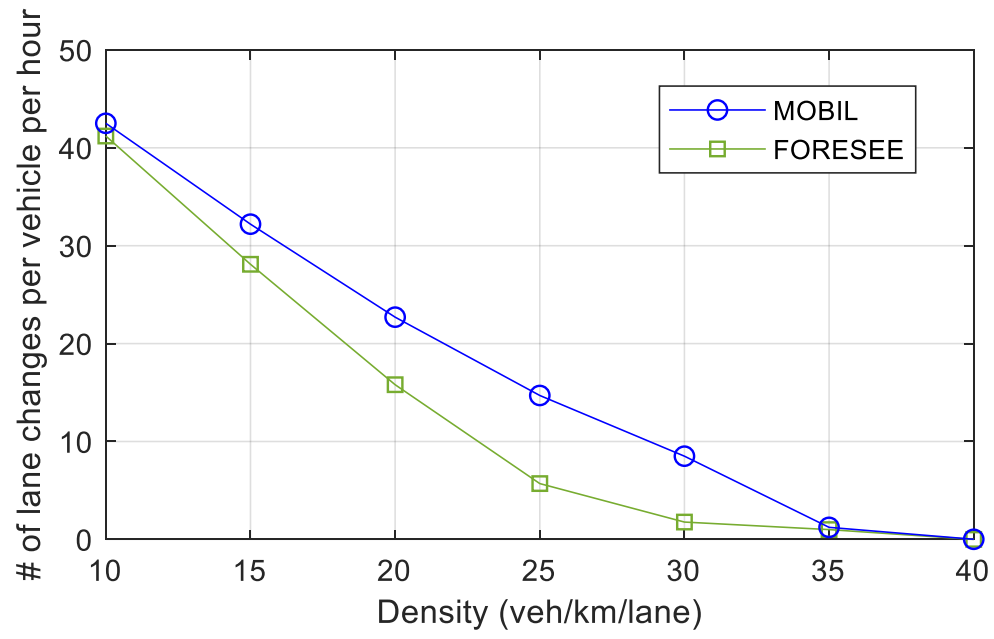

Fig. 19. Average number of lane changes normalized per vehicle and hour as a function of the density.

Cooperative mobility increases vehicle speeds with fewer lane changes, which in turn enhances driving comfort. Fig. 20 shows that these benefits are achieved without increasing energy consumption. This efficiency is attributed to:

- Better distribution of vehicles: a more balanced allocation of vehicles across lanes helps enhance traffic flow.
- More efficient driving: fewer lane changes result in less accelerations and decelerations, leading to smoother driving.
- Strategic truck routing: trucks, which consume approximately six times more energy than cars, are predominantly directed to the right lanes to enhance overall traffic speed.

The energy consumption is calculated by integrating the power request over time, and the power request is determined as follows [1]:

$$P(v,a) = \max\,[P_0 + v \cdot F(v,a), 0] \qquad (11)$$

where $v$ is the vehicle's speed, $a$ is the acceleration, $P_0$ is the idling power[2], and $F(v,a)$ is the driving resistance. $F(v,a)$ is equal to [1]:

$$F(v,a) = m \cdot a + c_R \cdot m \cdot g + \frac{1}{2}\gamma \cdot c_d \cdot A \cdot v^2 \qquad (12)$$

where $m$ denotes the mass of the vehicle, $g$ is the acceleration of gravity, $\gamma$ represents the air density, $A$ is the frontal surface area of the vehicle, and $c_r$ and $c_d$ are the coefficients for rolling resistance and air resistance, respectively. Table IV presents the values for these parameters, selected according to references [19][20]. Fig. 20 shows the average energy consumption across densities. The results show that FORESEE has a similar energy consumption as MOBIL. However, FORESEE results in higher average speeds in most of the traffic densities evaluated (Fig. 18), which corresponds to more efficient driving, since an increase in average speed for the same traffic demand generally implies higher energy consumption.

[2] We neglect $P_0$ as vehicles are in constant movement in the highway.

TABLE IV. PARAMETERS FOR ESTIMATING THE ENERGY CONSUMPTION

| | Trucks | Cars |
|---|---|---|
| Mass of the vehicle ($m$) | 29484 kg | 1500 kg |
| Frontal surface area ($A$) | 7.6 m² | 2.3 m² |
| Rolling resistance coefficient ($c_r$) | 0.006 | 0.015 |
| Air resistance coefficient ($c_d$) | 0.84 | 0.26 |
| Acceleration of gravity ($g$) | 9.8 m/s² | |
| Air density ($\gamma$) | 1.2 kg/m³ | |

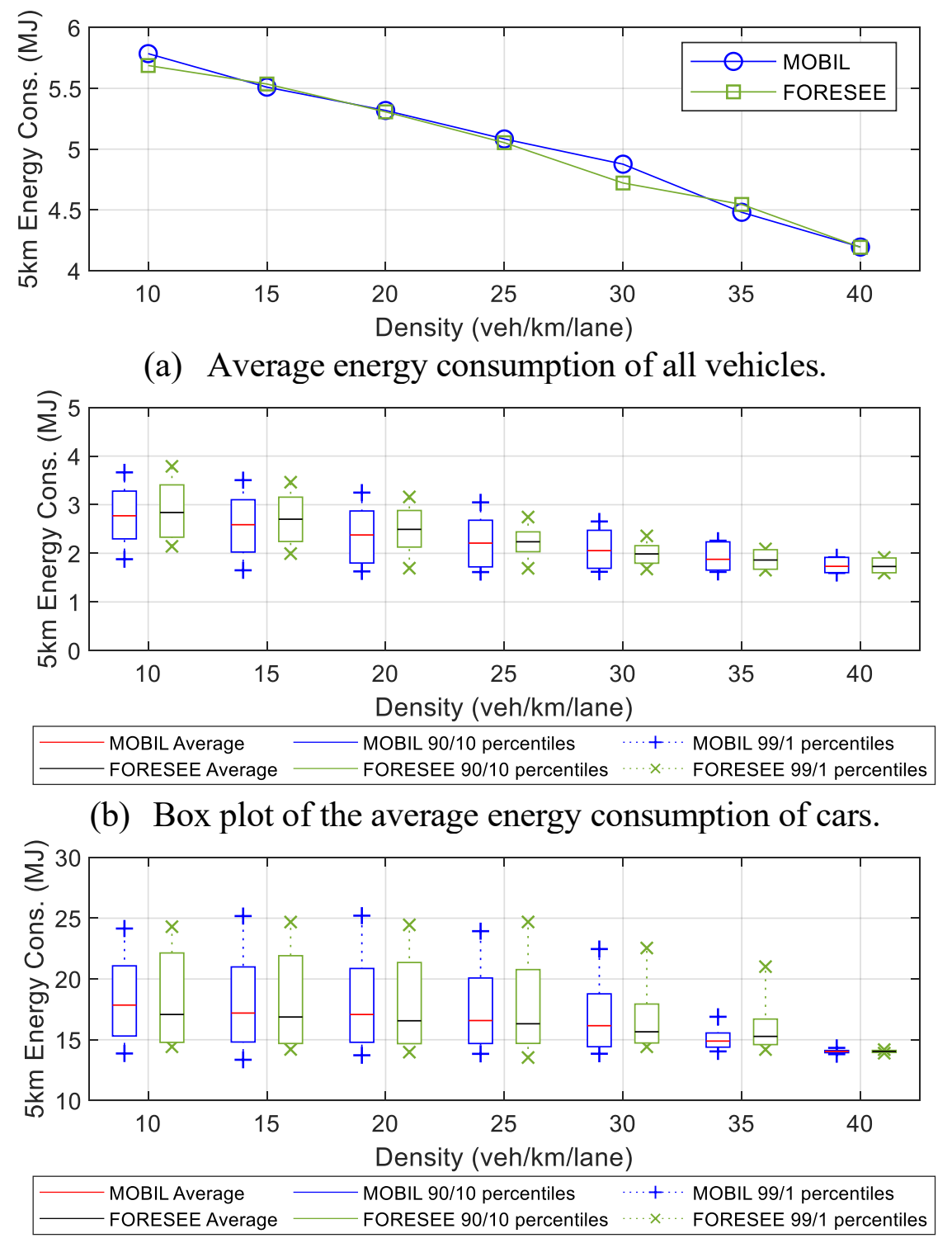

(a) Average energy consumption of all vehicles.

(b) Box plot of the average energy consumption of cars.

(c) Box plot of the average energy consumption of trucks.

Fig. 20. Energy consumption with cooperative (FORESEE) and non-cooperative (MOBIL) lane changes as a function of the density

The benefits of cooperative lane changes in enhancing traffic efficiency augment in the presence of road obstacles, which can significantly disrupt traffic flow. Fig. 21 illustrates the average speed of all vehicles in the scenario with a road obstacle, comparing cooperative (FORESEE) and non-cooperative (MOBIL) lane change strategies. The comparison between Fig. 18 and Fig. 21 demonstrates that cooperative lane changes yield greater improvements in the presence of road obstacles. This effect is most pronounced in scenarios with densities up to 30 vehicles/km/lane, where lane changes are feasible, and the ability to anticipate road conditions ahead can have a significant impact. In these conditions, cooperative lane changes can increase average speeds by 8% to 10%, resulting in a corresponding increase in road capacity by 942 vehicles per hour. It is important to note that the speed enhancements achieved with cooperative lane changes are particularly significant for vehicles in the lower 1st and 10th percentiles, which are the vehicles most affected by road obstacles. For instance, cooperative lane changes improve the average speed of the 1st percentile of cars between 53.4% and 93.3%, and the average speed of the 10th percentile between 11.6% and 30.5%. These gains underscore the capability of cooperative lane changes to enhance traffic efficiency by leveraging V2X communications to anticipate road conditions ahead.

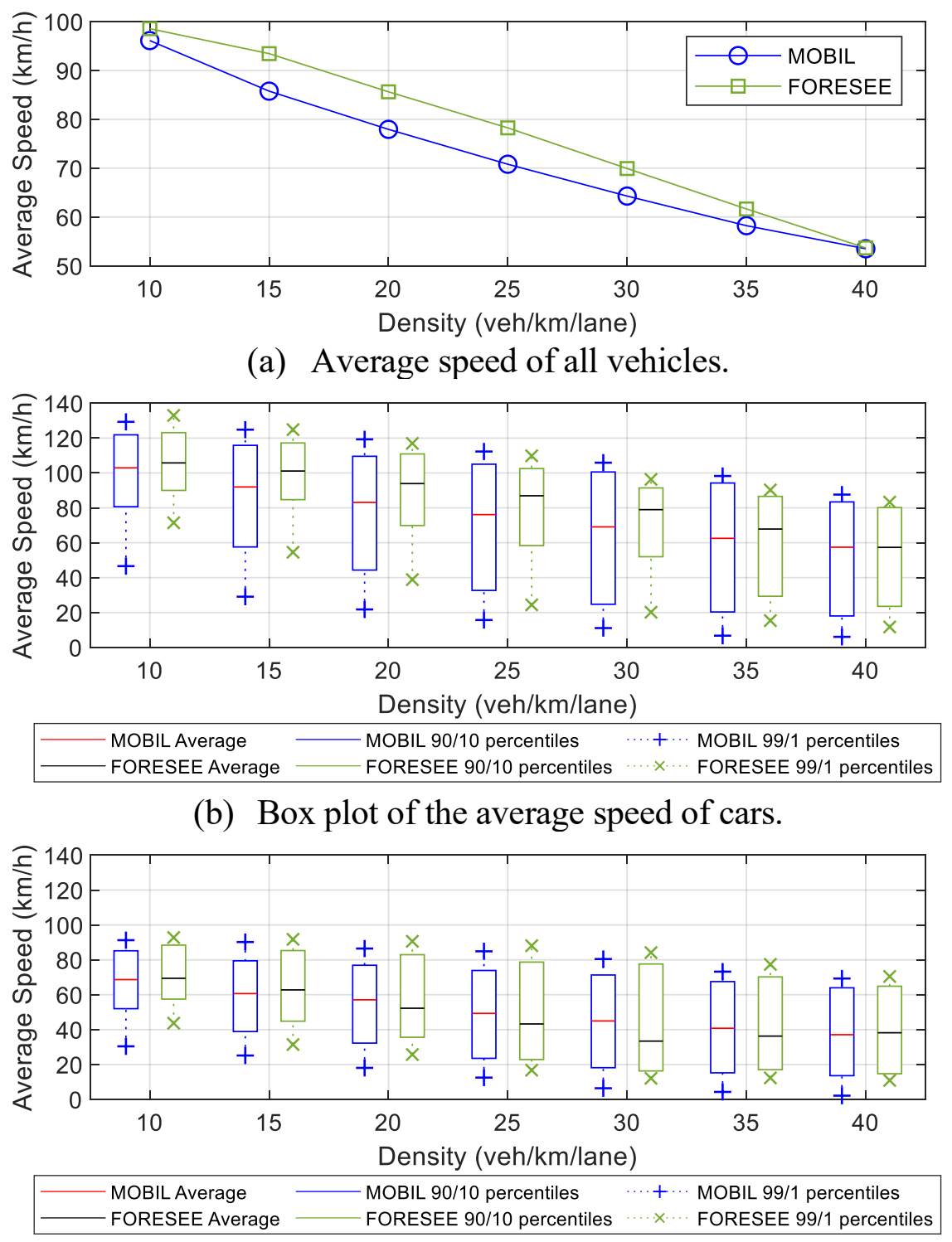

(a) Average speed of all vehicles.

(b) Box plot of the average speed of cars.

(c) Box plot of the average speed of trucks.

Fig. 21. Average speed with cooperative (FORESEE) and non-cooperative (MOBIL) lane changes as a function of the density in the scenario with a road obstacle .

## VI. ON THE POTENTIAL OF MANEUVER COORDINATION

Previous sections have demonstrated the potential of cooperative lane changes to enhance traffic efficiency through fewer and more effective lane changes. However, in many cases, lane changes are hindered by vehicles in adjacent lanes or insufficient gaps between vehicles. These lane changes could be possible if vehicles were able to coordinate their maneuvers. Through maneuver coordination, vehicles can exchange information about their intentions and coordinate their driving using V2X communications. As outlined in [21], maneuver coordination can be triggered when a vehicle intends to change lanes but cannot due to insufficient gap in the target lane. In the case of cooperative lane changes with FORESEE, a maneuver coordination would be triggered when the lane change is classified as 'wanted but not possible'. A lane change is classified as wanted when the incentive criterion is satisfied. A lane change is classified as possible only if the comfort criterion is also satisfied. Therefore, the cases computed as 'wanted but not possible' for a vehicle correspond to situations in which the incentive criterion is fulfilled, but the comfort criterion is not met. For example, consider an ego vehicle $V_{ego}$ that wants to change lanes, with vehicle $V_{bn}$ behind $V_{ego}$ in the target lane and insufficient gap for $V_{ego}$ to change lanes. In this case, if $V_{ego}$ wants to change lanes but it cannot do so, a maneuver coordination could involve a controlled deceleration of $V_{bn}$ to create enough gap for the lane change.

Fig. 22 illustrates the percentage of time that lane changes are wanted but cannot be executed with FORESEE in the scenario without obstacles. This percentage increases with traffic density because more vehicles want to change lanes to align with their desired speed but are unable to do so. Maneuver coordination could enable the execution of these desired lane changes that were previously deemed not possible in the current evaluation. High traffic densities inherently limit opportunities to find suitable gaps for desired lane changes. At very high densities (35 and 40 vehicles/km/lane), Fig. 22 shows that fewer vehicles want to change lanes. This is because the entire scenario slows down, and slow vehicles no longer have the incentive to change lanes.

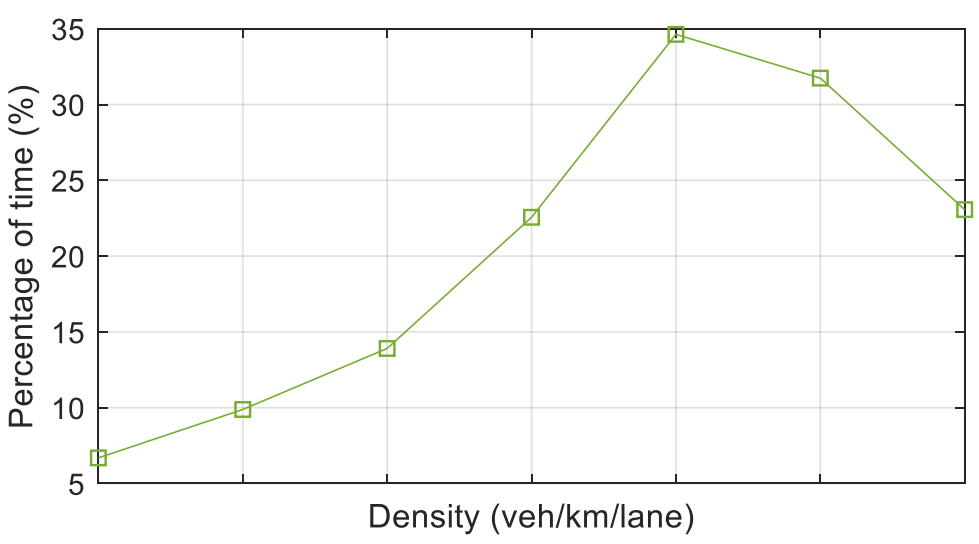

Fig. 22. Percentage of time when lane changes are wanted but not possible with FORESEE.

Fig. 23 highlights the potential of cooperative mobility and maneuver coordination in scenarios with obstacles. The figure shows the percentage of time that a lane change is desired but not possible for vehicles in the right lane, positioned behind the obstacle. Using the cooperative lane change model FORESEE, configured with $d_R$=500m, vehicles estimate the speed of each lane $S_l$. All vehicles within 500 meters of the road obstacle detect its presence through V2X communications and want to change lanes to bypass the obstacle, regardless of the traffic density. Vehicles located more than 500 meters away from the road obstacle will only want a lane change if the vehicles within their range $d_R$=500m are moving slowly. In this context, a vehicle 800 meters behind the obstacle is unlikely to want a lane change in low-density traffic because vehicles within its 500 m range are probably not affected by the obstacle which is still 300 meters away. However, as traffic density increases, the queue of vehicles behind the obstacle grows rapidly, impacting vehicles further back. In such high-density conditions, with FORESEE, a vehicle located 800 meters behind the obstacle can detect that vehicles within its $d_R$=500m are decelerating due to the obstacle's influence. Consequently, the percentage of time vehicles wish to change lanes but are unable to do so rises with density. These observations highlight the potential for maneuver coordination to facilitate lane changes that are desired but currently unfeasible before vehicles encounter road obstacles disrupting the traffic flow.

The analysis in this section reveals that many desired lane changes cannot be executed due to insufficient gaps in the target lane for the ego vehicle. This situation underscores the potential of cooperative driving to enhance further traffic efficiency through maneuver coordination. However, these coordinated maneuvers must be carefully designed and optimized to facilitate desired lane changes without causing traffic disruptions. This is particularly relevant as maneuver coordination may involve the deceleration of one or more vehicles to create the necessary gap, which can lead to traffic disturbances, especially in high-density conditions. Cooperative lane changes like FORESEE could compensate these decelerations as they help homogenize and improve traffic flow by promoting a better distribution of vehicles across lanes. Moreover, cooperative lane change models like FORESEE are essential for the proper design and testing of maneuver coordination strategies. Maneuver coordinations should be designed with the objective to improve the overall traffic flow (not just the immediate acceleration of an ego vehicle) in the short, medium, and ideally, long term. This requires leveraging look-ahead V2X information to understand the broader traffic context and to plan maneuvers accordingly. The effectiveness of maneuver coordinations is closely tied to the lane change model used. If V2X-based maneuver coordinations are tested with models that rely solely on local information, the lane change decisions may not align with the broader objectives of maneuver coordination, potentially undermining its impact on traffic efficiency.

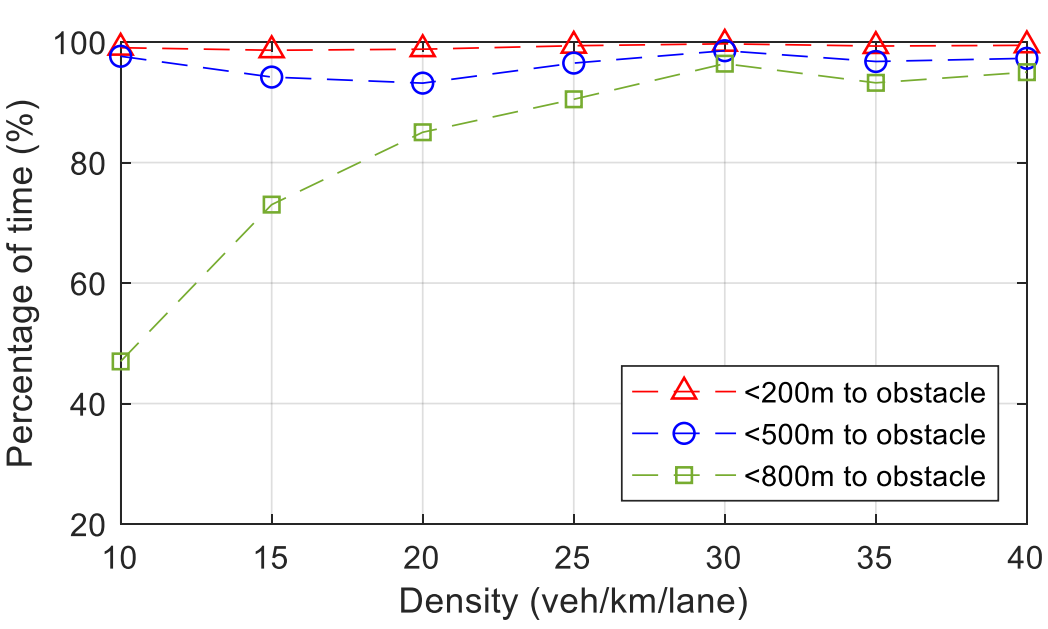

Fig. 23. Percentage of time when lane changes are wanted but not possible for vehicles behind the obstacle at various distances from the obstacle.

## VII. Conclusions

This paper presents FORESEE, a novel cooperative lane change model that leverages V2X data to anticipate traffic conditions and strategically organize lane changes, thereby enhancing overall traffic flow. FORESEE utilizes V2X data to estimate the speed of each lane and organizes vehicles into lanes based on their desired speeds. This approach contributes to the homogenization of traffic flow and helps mitigate disturbances caused by abrupt lane changes or braking maneuvers, which can be triggered by differences in speed between vehicles traveling in the same lane. Our evaluation demonstrates that cooperative lane changes using FORESEE improve the average speed of vehicles, road capacity, and energy efficiency. This is achieved through fewer but more effective lane changes. By strategically organizing lane changes with V2X look-ahead data, vehicles can maintain speeds closer to their desired speeds and experience fewer fluctuations in speed and acceleration, which, in turn, enhances driving comfort. Additionally, our study shows that cooperative lane changes can better manage road traffic disturbances, such as obstacles, by anticipating traffic conditions and organizing lane changes accordingly. This capability reduces the number of vehicles affected by disturbances and increases the average speed of vehicles.

The evaluation also highlights the potential for further enhancing traffic efficiency through maneuver coordination as desired lane changes can be impeded by vehicles in adjacent lanes or insufficient gaps, particularly as the traffic density increases. Maneuver coordination enables vehicles to share their intentions and coordinate their maneuvers using V2X communications. However, maneuver coordinations must be carefully designed and optimized to facilitate desired lane changes without causing traffic disruptions when one or more vehicles decelerate to create necessary gaps for coordinated maneuvers. Therefore, maneuver coordination should leverage V2X data to comprehend the broader traffic context and plan maneuvers that improve overall traffic flow. The effectiveness of maneuver coordination is closely tied to the organization and execution of lane changes. Thus, the availability of cooperative lane change models like FORESEE is important for the correct design and testing of maneuver coordination strategies.

## ACKNOWLEDGEMENTS

This work has been partially funded by MICIU/AEI/10.13039/501100011033 and “ERFD/EU” (PID2023-150308OB-I00), and the “European Union NextGenerationEU/PRTR” (TED2021-130436B-I00), and UMH’s Vicerrectorado de Investigación grants.